\PassOptionsToPackage{frozencache,cachedir=minted-cache}{minted}
\documentclass[format=acmsmall, screen=true, nonacm,pdftex,svgnames]{acmart}

\acmJournal{PACMPL}
\acmVolume{1}
\acmNumber{POPL}
\acmArticle{1}
\acmYear{2025}
\acmMonth{1}
\acmDOI{}
\startPage{1}

\setcopyright{none}

\newcommand{\ignore}[1]{{}}
\usepackage{xcolor}
\usepackage{colortbl}

\usepackage[utf8]{inputenc}

\usepackage{amsmath}
\usepackage{hyperref}

\usepackage{subcaption}
\usepackage{wrapfig}

\usepackage{xspace}
\usepackage{float}
\usepackage{balance}
\usepackage{algorithm}
\usepackage{algorithmic}
\usepackage{textcomp}

\usepackage{changepage}
\usepackage{array,etoolbox}

\preto\tabular{\setcounter{magicrownumbers}{0}}
\newcounter{magicrownumbers}

\usepackage{cleveref}

\usepackage{wrapfig}
\usepackage{caption}

\usepackage{longtable}
\usepackage{tabu}

\usepackage{wasysym}
\usepackage{mathpartir}

\usepackage{mathtools}
\usepackage{xfrac}

\usepackage[qm,braket]{qcircuit}

\usepackage{bbm}
\usepackage{booktabs}
\usepackage{subcaption}
\DeclareMathAlphabet{\mathpzc}{OT1}{pzc}{m}{it}

\newtheorem{theorem}{Theorem}[section]

\usepackage{stackengine}
\usepackage{upgreek}

\newcommand{\cif}[3]{\texttt{if}~{#1}~\cn{then}~{#2}~\cn{else}~#3}

\colorlet{kwd}{black!80!green}
\definecolor{spec1}{RGB}{78, 131, 162}
\definecolor{spec0}{RGB}{66, 102, 136}
\definecolor{lespec}{RGB}{30, 80, 180}
\colorlet{spec}{lespec}
\colorlet{auto}{lespec!35!lightgray}
\colorlet{stack}{magenta}

\newcommand{\sqir}{SQIR\xspace}

\newcommand{\qwire}{{Qwire}\xspace}
\newcommand{\qbricks}{{QBricks}\xspace}

\newcommand{\voqc}{\textsc{voqc}\xspace}

\newcommand{\myparagraph}[1]{\noindent\paragraph{\textbf{#1}}}

\usepackage{tikz}

\newcommand{\dom}[1]{\ensuremath{\mathtt{dom}(#1)}}

\newcommand{\ones}{\mathbb{1}}
\newcommand{\zeros}{\mathbb{0}}

\newcommand{\slen}[1]{|#1|}

\newcommand{\CC}{\ensuremath{\mathbb{C}}\xspace}

\usepackage{pgf}

\usepackage{tikz}
\usetikzlibrary{%
  arrows,%
  shapes.misc,%
  shapes.geometric,
  shapes.arrows,%
  shapes.callouts,
  shapes.gates.logic.US,
  chains,%
  matrix,%
  positioning,%
  scopes,%
  decorations.pathmorphing,%
  decorations.text,
  decorations.pathreplacing,
  shadows,%
  automata,
  fit, calc, arrows.meta
}

\tikzset{ machine/.style={
    rectangle,
    minimum width=25mm,
    minimum height=18mm,
    text width=24mm,
    align=center,
    very thick,
    draw=black,
    color=black,
    fill=white,
  }
}

\DeclarePairedDelimiter\abs{\lvert}{\rvert}
\DeclarePairedDelimiter\norm{\lVert}{\rVert}

\makeatletter
\let\oldabs\abs
\def\abs{\@ifstar{\oldabs}{\oldabs*}}
\let\oldnorm\norm
\def\norm{\@ifstar{\oldnorm}{\oldnorm*}}
\makeatother

\makeatletter
\DeclareRobustCommand{\vardivision}{%
  \mathbin{\mathpalette\@vardivision\relax}%
}
\newcommand{\@vardivision}[2]{%
  \reflectbox{$\m@th\smallsetminus$}%
}
\makeatother

\usepackage{booktabs}

\usepackage{alltt}
\usepackage{listings,lstcoq}
\definecolor{ltblue}{rgb}{0,0.4,0.4}
\definecolor{dkblue}{rgb}{0,0.1,0.6}
\definecolor{dkgreen}{rgb}{0,0.35,0}
\definecolor{dkviolet}{rgb}{0.3,0,0.5}
\definecolor{dkred}{rgb}{0.5,0,0}
\usepackage[export]{adjustbox}
\usepackage{subcaption}

\newcommand{\code}[1]{\lstinline|#1|}

 \newcommand{\rulelab}[1]{{\small \textsc{#1}}}

\DeclareMathOperator*{\Msum}{\text{\raisebox{0.25ex}{\scalebox{0.8}{$\sum$}}}}

\newcommand{\tjudge}[4]{#1\vdash_{#2} #3 : #4}
\newcommand{\tcompile}[5]{#1\vdash_{#2} #3 : #4\triangleright #5}

\newcommand{\pau}{\mathpzc{P}}

\makeatletter
\def\arcr{\@arraycr}
\makeatother

\newcommand{\sapp}[2]{#1\,#2}

\newcommand{\uapp}[2]{#1\, #2}

\newcommand{\gfun}{\mathpzc{G}}
\newcommand{\cfun}{\mathpzc{C}}
\newcommand{\rfun}{\mathpzc{R}}
\newcommand{\mnode}{\mathpzc{V}}
\newcommand{\medge}{\mathpzc{E}}
\newcommand{\dfun}{\mathpzc{D}}
\newcommand{\sfun}{\mathpzc{S}}
\newcommand{\ufun}{\mathpzc{U}}

\newcommand{\rset}[2]{\{#1\}^{\sim #2}}

\newcommand{\zero}{\mathpzc{O}}

\newcommand{\sdag}[1]{#1^{\dag}}

\newcommand{\sminus}{\texttt{-}}
\newcommand{\splus}{\texttt{+}}
\newcommand{\eexp}[1]{\cn{exp}(\sminus i #1)}

\newcommand{\qsnd}{$\textsc{MuQ}$\xspace}
\newcommand{\enq}{\textsc{EnQ}\xspace}
\newcommand{\enqp}{\textsc{EnQ}$^+$\xspace}

\newcommand{\qbind}[2]{\partial {#1} .\,#2\xspace}
\newcommand{\qbinds}[3]{\partial {#1} \in #2.\,#3\xspace}

\newcommand{\cn}[1]{\texttt{#1}}

\newcommand{\ife}[3]{\ensuremath{\cn{if}\ {#1} \ \cn{then}\ {#2} \ \cn{else}\ {#3}}\xspace}

\newcommand{\denote}[1]{\llbracket #1 \rrbracket\xspace}
\newcommand{\dabs}[1]{|\!| #1 |\!|}

\newcommand{\tob}[1]{\llbracket #1 \rrbracket}

\newcommand{\case}[3]{\ensuremath{\texttt{case}\ #1 \ \rhd \ 0: {#2}, \dots, n-1: {#3}}\xspace}

\usepackage{etoolbox}
\newtoggle{comments}
\toggletrue{comments}
\usepackage[normalem]{ulem}
\usepackage{minted}
\newcommand{\hask}[1]{\mintinline{haskell}{#1}}

\iftoggle{comments}{
  \newcommand{\fixme}[1]{\textbf{\textcolor{red}{[ Fixme: #1]}}}
  \newcommand{\todo}[1]{\textbf{\textcolor{green}{[ TODO: #1 ]}}}
  \newcommand{\mwh}[1]{\textbf{\textcolor{red}{[ Mike: #1 ]}}}
  
  \newcommand{\khh}[1]{\textbf{\textcolor{orange}{[ Kesha: #1 ]}}}
  \newcommand{\shh}[1]{\textbf{\textcolor{purple}{[ Shih-Han: #1 ]}}}
  \newcommand{\liyi}[1]{\textbf{\textcolor{orange}{[ Liyi: #1 ]}}}

 \newcommand{\chand}[1]{\textbf{\textcolor{blue}{[ Chandeepa: #1 ]}}}
  \newcommand{\oth}[2]{\textbf{\textcolor{red}{[ #1: #2 ]}}}
  \newcommand{\xwu}[1]{\textbf{\textcolor{purple}{[ Xiaodi: #1 ]}}}
  
  \newcommand{\ynote}[1]{\textbf{\textcolor{magenta}{[ Yi: #1 ]}}}

  \colorlet{MZ}{violet!80!pink}

  \newcommand{\mzr}[1]{{\color{MZ}{#1}}}
  \newcommand{\was}[1]{}
  
  \NewCommandCopy{\Creff}{\Cref}
  \renewcommand{\Cref}[1]{\mbox{\Creff{#1}}}

  \usepackage[inline]{enumitem}
  \colorlet{LC}{cyan!31!teal}

  \usepackage[inline]{enumitem}
}{
  \newcommand{\fixme}[1]{}
  \newcommand{\todo}[1]{}
  \newcommand{\rnr}[1]{}
  \newcommand{\mwh}[1]{}
  \newcommand{\khh}[1]{}
  \newcommand{\liyi}[1]{}
  \newcommand{\shh}[1]{}
  \newcommand{\xwu}[1]{}
  \newcommand{\oth}[2]{}
  \newcommand{\mzr}[1]{}
  \newcommand{\chand}[1]{}

  \newcommand{\ynote}[1]{}
}

\newtoggle{submission}
\toggletrue{submission}

\iftoggle{submission}{
  
}{
  
}

\usepackage{bbold}
\usepackage{minted}

\begin{document}

\title{Constrained Adiabatic Quantum Programming}

\author{Liyi Li}
\email{liyili2@iastate.edu}
\author{Mahir Anand}
\email{mahir@iastate.edu}
\affiliation{%
  \institution{Iowa State University}
  \country{USA}
}

\author{James Bryan Graves}
\email{gravjabr@iu.edu}
\author{Amr Sabry}
\email{sabry@iu.edu}
\affiliation{%
  \institution{Indiana University}
  \country{USA}
}

\def\titlerunning{Constrained Adiabatic Quantum Programming}

\begin{abstract}
 
\title{Constrained Adiabatic Quantum Programming}

To find a $k$-clique in a graph, adiabatic quantum computation (AQC) evolves a quantum system from the minimum of an initial cost function to the minimum of a problem cost function that penalizes non-cliques. Existing AQC languages fix the initial cost function, so the search ranges over all $2^n$ subsets of the $n$ vertices, although only the subsets of size $k$ can be answers. We present \enqp{}, an AQC language in which the programmer writes both cost functions. Data constraints define the problem cost function. Relational constraints, a new construct, declare the moves the search may make between configurations and define the initial cost function. For the clique, a single relational constraint, which exchanges a chosen vertex with an unchosen one, confines the search to the subsets of size $k$. \enqp{} thereby turns constrained quantum annealing, which physicists set up by hand for individual problems, into a programming construct. We give \enqp{} a semantics in many-world configuration (MWC) graphs, whose nodes are configurations and whose edges are moves, and derive from it a step-wise semantics of adiabatic evolution. We prove that a run stays within the component of the MWC graph that contains its initial state, and that a run from an approximate initial state still reaches the answer, with probability equal to its squared overlap with the exact one. We compile \enqp{} to quantum circuits, prove the compilation correct with respect to the semantics, and use \enqp{} to program the Three-Dimensional Matching, Hamiltonian Cycle, and Traveling Salesperson problems.

\end{abstract}

\maketitle

\section{Introduction}\label{sec:intros}

Quantum computers aim to solve highly complex problems by leveraging quantum superposition, which enables substantially faster algorithms, such as Shor's algorithm, than classical algorithms.
A key issue in quantum computation is a high-level program abstraction, beyond the current digital circuit based model.
Adiabatic quantum computation (AQC) is a model for optimization problems~\cite{farhi2000quantum} and has been proven equivalent to the digital circuit model~\cite{1366223}.
We intend to develop a high-level programming paradigm, \enqp, based on AQC by connecting it with classical constraint programming.

An AQC problem is presented as a cost function over candidate solutions whose minimum is the answer; physically, this cost function is a \emph{Hamiltonian} and its minimum a \emph{ground state}, but we use the programming terminology throughout. The minimum cannot be prepared directly, since that is the problem itself, so AQC reaches it through a pair of cost functions. It initializes the system at the minimum of an \emph{initial} cost function $\hat{H}_i$, one whose minimum is easy to prepare, and then deforms the cost function slowly along $(1-t)\hat{H}_i + t\hat{H}_f$ into the \emph{problem} cost function $\hat{H}_f$, whose minimum is the answer. The adiabatic theorem guarantees that, deformed slowly enough, the system stays at the minimum the whole way and hence ends at the minimum of $\hat{H}_f$. How slow ``slowly enough'' must be is governed by the \emph{spectral gap}, the energy separating the minimum from the next configuration during the deformation; the smaller the gap, the longer the run.

\subsection{A Motivation Example: Clique Finding}

To see what is at stake, consider the standard Hamiltonian encoding of the
$k$-clique graph problem~\cite{Lucas_2014}. It puts a binary variable $x_v \in \{0,1\}$ on every
vertex of a graph $G = (V,E)$, so a configuration is a $0$/$1$ labeling of all
vertices; its cost is computed by the formula
\[
\Big(k - \sum_{v\in V} x_v\Big)^2 \;+\; \Big(\tfrac{k(k-1)}{2} - \sum_{(u,v)\in E} x_u x_v\Big)^2 .
\]
The first term grows as the number of $1$s departs from $k$; the second grows as the $1$-valued vertices miss the $\tfrac{k(k-1)}{2}$ edges of a clique. The cost is zero exactly on a $k$-clique. The search ranges over configurations with \emph{every} number of $1$s, which is clearly inefficient. Furthermore, a clique with the wrong count is penalized but not ruled out, so the run can still settle into a near-miss, such as a $(k\!-\!1)$-clique.

The AQC model has several issues.
First, the energy constraints above focus too much on developing a matrix form and include many compilation details.
Essentially, the $(\tfrac{k(k-1)}{2} - \sum_{(u,v)\in E} x_u x_v)^2$ simply suggests that one needs to look at every pair of configuration nodes in $G$ and examine if there is an edge connecting them; the sum and square operations exist because they are required by the compilation.
Second, the above expression defines only the $H_f$ part, while the initial Hamiltonian ($H_i$) is not clearly explained.
The initial state $\ket{\varphi}$ is required to be an eigenstate of the initial Hamiltonian $H_i$, but the model provides no additional information on how users can understand and program $H_i$. Third, a more essential issue is the interpretability of the relationship between the elements in the Hamiltonians and the elements in the initial quantum state.
For example, the binary variables $x_v$ can indeed be thought of as qubits in the quantum state, but it is unclear what happens if users want more than binary variables, what it might affect the quantum state to put two items $x_u x_v$ together, and what the hidden constraints are for the relations.

\begin{figure}[t]
    \centering
    \includegraphics[width=1\textwidth]{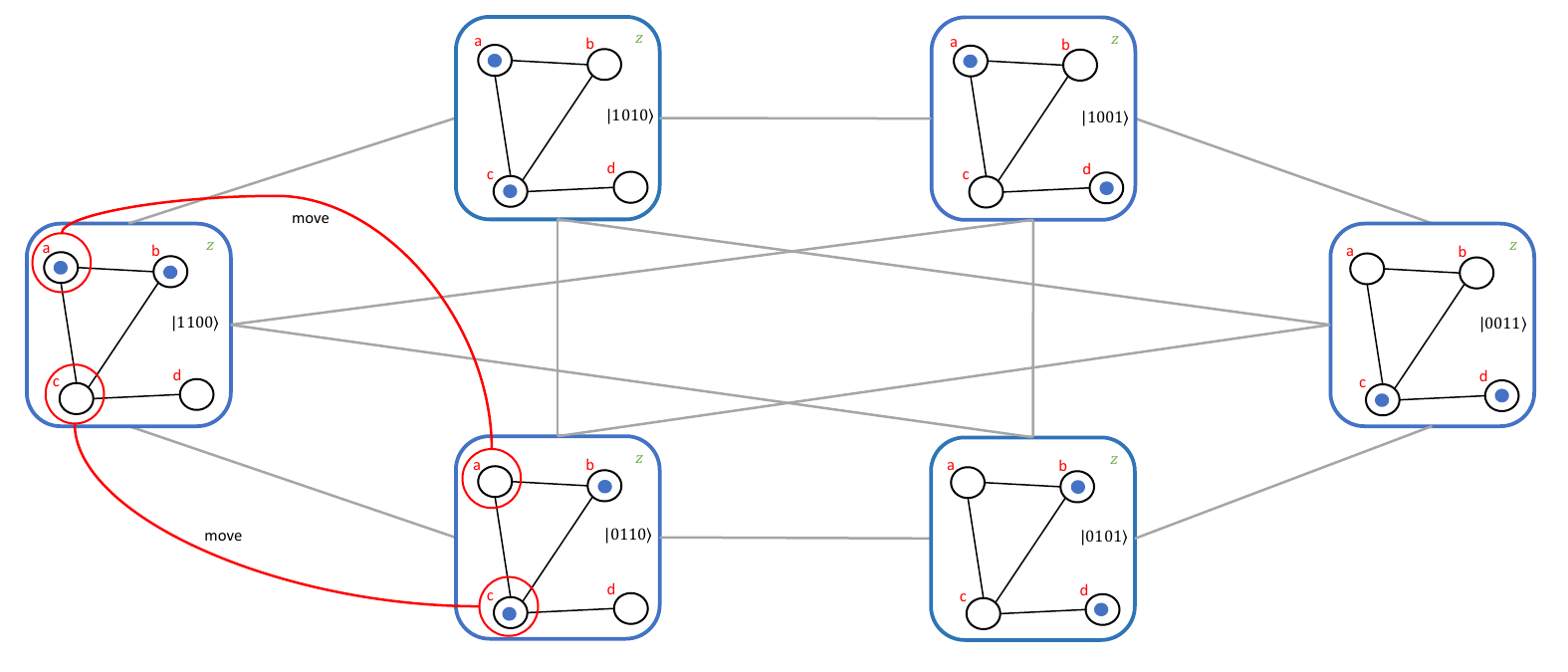}
    \caption{Example \emph{many-world configuration} graph for the state $\frac{1}{\sqrt{6}}(\ket{1100}+\ket{1010}+\ket{1001}+\ket{0011}+\ket{0101}+\ket{0110})$; $z=\frac{1}{\sqrt{6}}$. Two configurations are connected with a gray line only if they are exactly different by two data points.}
    \label{fig:exmwc-intro}
\end{figure}

Our first contribution is developing an interpretation of the AQC model based on a data structure rather than a pure quantum state.
We develop a many-world configuration graph for interpreting the quantum states used in the model.
\Cref{fig:exmwc-intro} shows an example many-world configuration (MWC) graph for a \emph{Dicke state} used in the clique-finding problem.
Assuming the data-structure graph in clique finding has four data-point vertices, with each data-point containing a field indicating its subgraph inclusion; e.g., blue vertices are included in the subgraph when considering a clique.

The MWC graph can be viewed as a high-level abstraction of a quantum state $\ket{\varphi}$, where we map each basis-ket state in $\ket{\varphi}$ to a configuration in the graph and each data-point vertex is a bit in the basis-ket, e.g., the left configuration represents $\frac{1}{\sqrt{6}}\ket{1100}$, so we mark the blue vertices for the $a$ and $b$ positions.

MWC graphs support AQC constraints: initial Hamiltonians ($H_i$) define the \emph{relational constraint} relating edges in the MWC graph, and final Hamiltonians ($H_f$) define the \emph{data constraint} within a configuration.
In the clique finding example, a data constraint is defined for every data-point pair in a configuration, e.g., for every data-point pair ($(a,b), (a,c), (a,d), (b,c), (b,d), (c,d)$), we give an energy penalty if the pair is included in the subgraph and there is no edge between them; e.g., there is an energy penalty for $(b,d)$ at the configuration $\ket{0101}$.

Relational constraints are defined among MWC graph edges connecting two configurations. A valid edge depends on the specific relational constraint being defined.
In clique finding, we look at data-point pairs; thus, an edge connects two configurations that differ in exactly two data points, e.g., in the edge connecting the $\ket{1100}$ and $\ket{0110}$ configurations, the $a$ and $c$ data points differ. In \enqp, we can constrain this relation by swapping the choice values between the two data points in the two configurations.

As mentioned above, a typical requirement is that the program's initial state is the ground eigenstate of the relational constraint (initial Hamiltonian), but many initial states are hard to prepare.
A key \enqp contribution examines the case where we have an ideal relational constraint but an easy-to-prepare initial state that is not the relational constraint eigenstate, detailed in \Cref{thm:aqcorder1}.

Another major contribution is a certified compiler that relies on developing a dynamic semantics for AQC, with a theorem that shows the final outcome, though it omits many details in the step-wise transition semantics.
With the added probabilistic feature and without dynamic semantics, users who compile an AQC problem to quantum computers may struggle to tell whether the issue is a compilation problem or simply that they need to wait a long time. Even if a good outcome is produced eventually, it is hard to evaluate the compilation's effectiveness.
\enqp develops the first dynamic semantics for AQC based on a widely used setting \cite{10.1145/3747521,10.5555/2011430.2011431,Hen2016constrained,Hadfield2019qaoa,Fuchs2022mixers,QiskitQUBO,DWaveCQM,OpenQAOAMixers} and uses the semantics to certify our AQC compilation from \enqp programs to quantum computers.
Using the certified compiler, we uncover a long-standing issue in Hamiltonian-simulation-based AQC compilation.

\subsection{Contributions and Roadmap}

\enqp is based on AQC and builds on prior work. The detailed AQC background and the difference between \enqp and previous works in physics are given in \Cref{sec:physics}.
We identify the following contributions.

\begin{itemize}

\item We develop the MWC graph as an interpretation of quantum states as high-level data structures, serving as the state representation for \enqp; details are in \Cref{sec:staterepresentation}.

\item We extend \enq \cite{10.1145/3747521} with additional relational constraints, permitting users to constrain the initial starting points when solving a problem, detailed in \Cref{sec:syntax}.

\item We define the formal constraint semantics and derive the dynamic transition semantics for \enqp programs. To the best of our knowledge, this is the first step-wise transition semantics for AQC for the specific \enqp setting, detailed in \Cref{sec:sema}.

\item Using the formal semantics, we prove a good approximation theorem, examining the situation where the initial state is not the eigenstate of a relational constraint, detailed in \Cref{sec:sema}.

\item Using the type system \Cref{sec:typing}, we develop a certified compiler that compiles \enqp programs to quantum computers and uncover an issue in the previous AQC compilation (\Cref{sec:discussion}).

\end{itemize}

\section{Overview through a Worked Example: Finding a 3-Clique}
\label{sec:example}

\begin{wrapfigure}{r}{3.5cm}
  \includegraphics[width=0.27\textwidth]{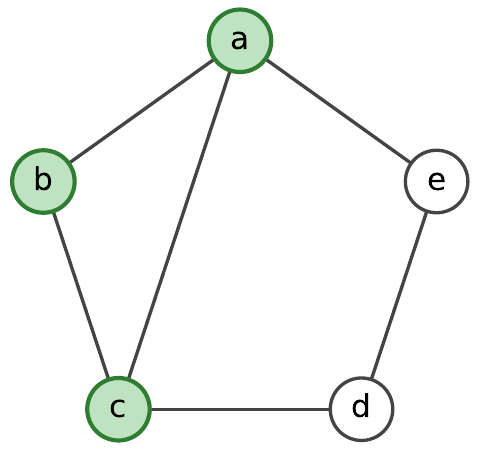}
  \caption{Our graph. Its only triangle, $\{a,b,c\}$, is shaded.}
            \label{fig:exgraph}
          \end{wrapfigure}

This section provides an \enqp overview based on an example of finding a $3$-clique in the five-vertex graph of \Cref{fig:exgraph}, whose edges are $ab$, $bc$, $cd$, $de$, $ea$, and the chord $ac$. The graph has exactly one triangle, $\{a,b,c\}$.

\myparagraph{Background}
\label{sec:aqcbackground}
Adiabatic Quantum Computation (AQC) is a model of quantum computation that solves optimization problems by evolving a quantum system, defined by a Hamiltonian $\hat{H}(t)=(1-t)\hat{H}_i + t \hat{H}_f$, slowly from an initial Hamiltonian $\hat{H}_i$ with a known ground state to a final Hamiltonian $\hat{H}_f$ whose ground state encodes the solution. The evolution adjusts $t$ slowly every step from $0$ to $1$. By maintaining the system in its lowest energy state throughout the process, AQC leverages the adiabatic theorem to find optimal solutions efficiently. The first line below shows the Hamiltonian $H$ that varies with time $t$ from $H_i$ to $H_f$. The exact evolution of this Hamiltonian would be given by the Schr\"odinger equation on the right.
A typical way of approximating the AQC solution uses \emph{Hamiltonian simulation} consisting of the sequence on the second line, where each step generates a quantum operation via a matrix exponential operation ($\eexp{\sapp{\hat{H}(t)}{\Delta t}}$) for a small time period $\Delta t$.
Note that a Hamiltonian is a Hermitian matrix (Hermitian $H = \sdag{H}$) and that Hamiltonian simulation ($\eexp{\hat{H} t}$) generates a unitary matrix that can be implemented as a quantum circuit. Therefore, the AQC procedure can be understood as starting with an initial quantum state $\ket{\varphi(0)}$ and applying a sequence of unitary quantum operations, each of which is of the form $\eexp{\sapp{\hat{H}(t)}{\Delta t}}$ for a different $t\in\{\Delta t, 2\Delta t, 3 \Delta t, ...,T\}$.

{\small
\begin{center}
$
\begin{array}{l}
\hat{H}(t)=(1-\frac{t}{T})\hat{H}_i + \frac{t}{T}\hat{H}_f
\qquad
\qquad
i\frac{d}{dt}\ket{\varphi(t)}=\hat{H}(t)\ket{\varphi(t)}
\\[0.4em]
\ket{\varphi(T)}\approx \eexp{\sapp{\hat{H}(T)}{\Delta t}}...\eexp{\sapp{\hat{H}(2\Delta t)}{\Delta t}}\eexp{\sapp{\hat{H}(\Delta t)}{\Delta t}}\ket{\varphi(0)}
\end{array}
$
\end{center}
}

\begin{figure}[t]
    \centering
    \includegraphics[width=0.46\textwidth]{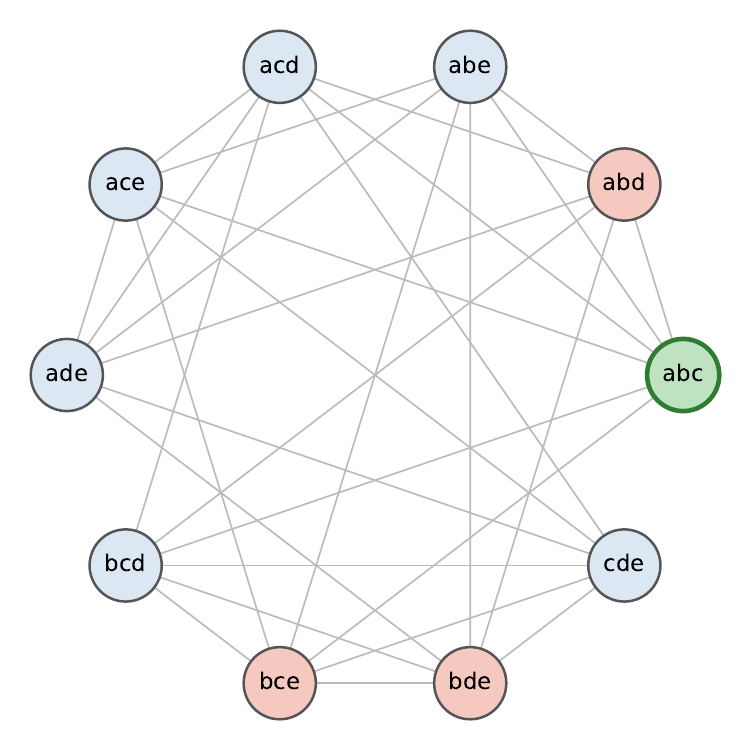}
    \caption{The \emph{many-world configuration} graph: the ten candidates, joined
    wherever a single swap turns one into another. Green is the clique
    ($\hat{H}_f=0$), blue costs $1$, red costs $2$.}
    \label{fig:exmwc}
\end{figure}

\myparagraph{Understand Constraints based on MWC Graphs}\label{sec:constraintintro}
We first show the \enqp interpretation of data constraints on the MWC graphs, corresponding to the final Hamiltonian ($\hat{H}_f$) in an AQC pair.
\Cref{fig:exmwc} is an example MWC graph for the clique-finding problem in \Cref{fig:exgraph}, based on an initial-and-data-constraint setting; discussed shortly below.
In clique finding, each vertex has a binary choice value, denoted $x_v\in \{\cn{Yes},\cn{No}\}$, where \cn{Yes} means the vertex is included in the clique. A \emph{configuration} is one such labeling of all five vertices, and there can be $2^5 = 32$ of them.
For example, \Cref{tab:exconfigs} marks the vertices $a$, $b$, and $c$ to be $\cn{Yes}$, and the others to be $\cn{No}$.
In \enqp, we define the data constraint to be as follows.

{
\[
\qbind{x}{\qbind{y}{\ife{\neg\cn{edge}(x,y)\wedge\cn{Yes}(y)\wedge\cn{Yes}(x)}{10}{0}}}
\]
}

Given two vertices, $x$ and $y$, in a graph, if they are considered in the subgraph ($\cn{Yes}(y)\wedge\cn{Yes}(x)$) and there is no edge between them, we assign an energy loss to them; otherwise, the loss is $0$, e.g., \Cref{tab:exconfigs}; only $\{a,b,c\}$ is minimal; the others each pay for one or two of the missing edges they contain. We apply the constraint to every vertex pair in the graph and sum the energy scores. If we find a configuration $\phi$ with score $0$, the \cn{Yes} vertices in $\phi$ form a clique.
As in the example above, a data constraint on a MWC graph examines the relations among data points inside a configuration node.

One issue is setting the energy scores. For example, the energy scores for the above data constraint can be set to different values; such differences might affect the evolution time but not the final result. When dealing with an AQC problem, users might not know the exact energy scores to set but can identify a direction to set the energy scores, because setting the exact energy scores depends on many factors, such as compilation and quantum hardware.
For example, in the above example, it might be more precise to do the following.

{
\[
v \triangleq \cn{L} \mid \cn{O}
\qquad
\qbind{x}{\qbind{y}{\ife{\neg\cn{edge}(x,y)\wedge\cn{Yes}(y)\wedge\cn{Yes}(x)}{\cn{L}}{\cn{O}}}}
\]
}

Here, we set the energy loss to \cn{L}, meaning there is some energy loss, but we leave the exact values to the compiler. Similarly, \cn{O} means a lowest energy state, not necessarily to be $0$.
In \enqp, we use an abstract data type and allow both exact and relative energy scores. When compiling the former, we encode exact energy values; for the latter, we use an optimal energy-score calculation.

A MWC graph is essentially a reinterpretation of a quantum state, serving as the data domain to which the \enqp constraints apply.
In a many-worlds interpretation, a quantum state can be viewed as a set of possible "events", which is expressed as a linear sum of basis-kets; i.e., the basis-vector of a basis-ket is a possible world, and its amplitude represents the probability of the world happening.
MWC follows the many-worlds interpretation, where each basis-ket is a possible configuration, and the whole quantum state is the collection of all possible configurations.

\begin{table}[t]
\centering\small
\begin{tabular}{lc@{\hspace{3em}}lc}
\toprule
candidate & $\hat{H}_f$ & candidate & $\hat{H}_f$\\
\midrule
$\{a,b,c\}$ & $\mathbf{0}$ & $\{a,d,e\}$ & $1$\\
$\{a,b,d\}$ & $2$ & $\{b,c,d\}$ & $1$\\
$\{a,b,e\}$ & $1$ & $\{b,c,e\}$ & $2$\\
$\{a,c,d\}$ & $1$ & $\{b,d,e\}$ & $2$\\
$\{a,c,e\}$ & $1$ & $\{c,d,e\}$ & $1$\\
\bottomrule
\end{tabular}
\caption{The ten three-vertex candidates and their cost
$\hat{H}_f = x_a x_d + x_b x_d + x_b x_e + x_c x_e$. Only $\{a,b,c\}$ costs $0$.}
\label{tab:exconfigs}
\end{table}

\myparagraph{Developing Relational Constraints.}
Beyond data constraints, we develop relational constraints based on MWC graph edges that connect data points across configurations.
The clique-finding example in \Cref{fig:exgraph} can have $32$ different configurations, while the MWC graph in \Cref{fig:exmwc} has 10 valid configurations, because we develop the following relational constraint to trim unqualified configurations.

{
\begin{equation}
\qbind{x}{\qbind{y}{\rset{x,y}{*}}}
\label{eq:swapham}
\end{equation}
}

Here, we examine pairs of configurations, connected by edges $(\phi_1,\phi_2)$.
The bindings $x$ and $y$ indicate that the constraint applies to two data points in $(\phi_1,\phi_2)$, e.g., in the configurations $\{a,c,d\}$ and $\{a,c,e\}$ of \Cref{fig:exmwc}, we examine the two data points $d$ and $e$ and relate the choices in the two data points.
The $*$ flag suggests that the symmetry relation is a swap, i.e., the choice in a data-point ($d$ or $e$) is swapped from one configuration ($\phi_1$) to the other ($\phi_2$).
In the MWC graph, two configurations are connected by an edge only if exactly one pair of data points satisfies the relational constraint, while the other data points remain the same across the configuration pair.
The swap constraint ensures that 1) only two data-points in a configuration pair admit a choice swap relation, and 2) the other data-points remain the same.
Thus, the number of \cn{Yes} choices is the same across different configurations.
The MWC graph \Cref{fig:exmwc} is initialized as each configuration having exactly $3$ choices; thus, it contains $\binom{5}{3} = 10$ configurations.
These ten configurations are the entire search space and are formalized using the \emph{many-world configuration} (MWC) graph of \Cref{fig:exmwc}: a node for each candidate, and an edge wherever a single swap turns one into another.

The above relational constraint defines the swapping relation, but other kinds of relational constraints may exist. For example, $\qbind{x}{\qbind{y}{\qbind{z}{\rset{x,y,z}{*}}}}$ defines a relational constraint describing a swap among three vertices in a configuration pair.

Another constraint is $\rset{x,y}{0}$, which defines a relation that moves in the same direction in a configuration pair.
Consider a Bell pair $\frac{1}{\sqrt{2}}(\ket{00}+\ket{11})$, which can be thought of as two qubits measuring into the same bits.
Viewing a Bell pair in \enqp can be thought of as a relational constraint that two qubits move in the same direction, i.e., the basis-vector set contains $\{\ket{00},\ket{11}\}$, where the two sites are always moving in the same direction, as $\ket{00} \leftrightarrow \ket{11}$.
An example of using such a constraint is in \Cref{sec:cases}.

One question is left: where does the search begin?
We permit users to define an initial state, interpreted as a MWC graph, as the starting point, so we start somewhere easy and well defined and let the evolution carry us the rest of the way. In the above example, the natural choice treats all ten candidates alike: the uniform superposition over them, known as the Dicke state; it minimizes precisely the relational constraint we wrote, so the programmer gets a sound starting point for free.

The issue is that an arbitrary initial state might be hard to prepare; for example, a quantum superposition of all permutations might be hard to prepare.
We want to examine the possibility that an initial state is not the eigenstate of the relational constraint (initial Hamiltonian).
For example, instead of having a MWC graph with 10 configurations, as in \Cref{fig:exmwc}, what happens if we use a full MWC graph with 32 configuration nodes, while keeping the relational constraint in \Cref{eq:swapham}?

The good approximation theorem in \Cref{thm:aqcorder1} shows that if we can partition all configurations into relevant and irrelevant sets, with the condition that the two sets are clearly separated, the AQC evolution procedure does not affect the configurations in the irrelevant set, i.e., the amplitudes of configurations are not affected during computation, while the configurations in the relevant set are evolved in a trend similar to the case when the irrelevant set is empty; therefore, the configuration with the highest probability at the end is still the solution for the problem.

\myparagraph{Dynamic Semantics and Optimal Compilation}
\label{sec:gap}
The above interpretation of a constraint provides the constraint semantics, which constrains the user-desired system outcome, while we need dynamic semantics as a standard for guaranteeing compilation correctness. We can derive such a dynamic semantics from the constraint semantics, as shown in \Cref{def:dynsem}.

For compilation, the key is an enumeration process that recognizes a quantum state change, relying on four operations, $\zeros$, $\ones$, $\uparrow$, and $\downarrow$, with the following operational behavior.

{\small
\[
\zeros\ket{0} := \ket{0}
\quad
\zeros\ket{1} := \zero
\quad
\ones\ket{0} := \zero
\quad
\ones\ket{1} := \ket{1}
\quad
\uparrow\ket{0} := \ket{1}
\quad
\uparrow\ket{1} := \zero
\quad
\downarrow\ket{0} := \zero
\quad
\downarrow\ket{1} := \ket{0}
\]
}

To compile data constraints, we use the $\zeros$ and $\ones$ operations.
In clique finding, we essentially enumerate \cn{Yes} and \cn{No} as $\ket{1}$ and $\ket{0}$, respectively.
For every configuration, we generate every possible pair ($x$ and $y$) of data points, and interpret the two choices in the data-constraint $f$, where $f(\cn{choice}(x),\cn{choice}(y))$ is the energy loss score for $(x,y)$. We then summarize all the energy scores for each possible pair as the final energy loss for the configuration.
Essentially, we can view $f$ as a function that interprets the choices for a data-point pair. For example, in the $\{a,c,d\}$ configuration of \Cref{fig:exmwc}, we interpret the pairs $(a,b)$, $(a,c)$, $(c,d)$ with different choice combination. To interpret $(a,b)$, we take its choice value, $(1,0)$, and input it into the data constraint to query the energy score.

Recall that the choices are represented as basis-vectors in a quantum state, e.g., $(1,0)$ is represented as $\ket{1}\ket{0}$.
A quantum state is not observable, so we use the operators $\zeros$ and $\ones$ to represent $\ket{0}$ and $\ket{1}$, respectively.
For example, $(1,0)$ is represented as $\ones \otimes \zeros$ and we assign the energy score as the amplitude, $f(1,0) \cdot \ones \otimes \zeros$.
Finally, we sum all the generated operations, e.g., $f(1,0) \cdot \ones \otimes \zeros$, together via a linear sum.
In \enqp, we also instantiate the abstract energy scores with actual values to optimize the process.

Similarly, we can compile a relational constraint as a combination of $\uparrow$ and $\downarrow$.
For example, to compile the relational constraint above, we consider two configurations connected by an edge $(\phi_1,\phi_2)$.
We then compare the same name (\cn{id}) data-points across $\phi_1$ and $\phi_2$, having different choices, such as $d$ and $e$ in the pairs $\{a,c,d\}$ and $\{a,c,e\}$.
A choice changes $1 \to 0$ and $0 \to 1$ across the two data points, yielding a $\downarrow$ and $\uparrow$ operation, respectively, e.g., the pair $(d,e)$ is compiled to $\downarrow \otimes \uparrow+ \uparrow\otimes \downarrow$, where the linear sum ensures the symmetry property.
The above compilation compares a single configuration pair; to compile all configuration pairs, we use linear sums.

Apparently, the above procedure generates a linear sum of four operations, which is a syntactic sugar of Pauli operations. We then rewrite the operation into Pauli-string format $\sum_j \otimes_{k=0}^n \pau_k$ with $\pau \in \{\cn{X},\cn{Y},\cn{Z},\cn{I}\}$, via the rules below.

{\small
\[
\zeros\, := I + Z
\qquad
\ones \,:= I - Z
\qquad
\uparrow\, := X - iY
\qquad
\downarrow \,:= X + iY
\]
}

After producing a linear sum of Pauli strings, $\hat{H} = \sum_j \otimes_{k=0}^n \pau_k$, via distributive rewrites, we then utilize the approximation procedure in \Cref{sec:aqcbackground} to compute the AQC of the Pauli-string formula, while compiling the Hamiltonian simulation procedure via Trotterization.

While developing our certified compiler, we discovered a fault in previous AQC compilation papers; see \Cref{sec:discussion}.

\section{The \enqp Formalism}\label{sec:formalism}

This section discusses the syntax, semantics, type system, and meta-theorems for \enqp.
We use the running example of the clique finding problem (\Cref{fig:exgraph}) with the MWC graphs in \Cref{fig:exmwc}.

\subsection{State Representation}\label{sec:staterepresentation}

We discuss the state used in \enqp, built on an MWC graph, where each node contains a configuration for a user-defined problem.
We first define user-defined data structures below, having type $\gfun(\delta)$, assumed to contain data points with type $\delta$.

\begin{definition}[Data Structure and Configuration]\label{def:data-state}\rm
We define $G\in\gfun(\delta)$ to be a \emph{data structure}, containing data-points with type $\delta$.
Each data-point object must at least contain the fields of identity ($\cn{id}$), datum ($\cn{data}(\delta)$), and choice ($\cn{choice}(\delta)$).
We assume data ($\cn{data}(\delta)$) has an energy value type ($\rfun$) and a set of choices $\cn{choice}(\delta)=\zeta\in \cfun$ that turns a choice into an energy value. Given $\cfun$, a \emph{configuration} is a data structure $G$, with choice assignments ($\delta \to \cfun$) in $\cfun$ to every data-points in $G$.
\end{definition}

The example graph in \Cref{fig:exgraph} has $5$ data-points, $\{a,b,c,d,e\}$.
There is no need to use the data fields in clique finding, where we can define the data as $\forall x . \cn{data}(x)=0$, and choices for each node can be $\{\cn{no},\cn{yes}\}$.
The solution has the assignment $\{a\mapsto \cn{yes}, b\mapsto \cn{yes},c\mapsto \cn{yes},d\mapsto \cn{no},e\mapsto \cn{no}\}$.

We show the MWC graph definition below.

\begin{definition}[MWC Graph]\label{def:mwc-state}\rm
We define $\Phi=(\mnode,\medge)$ to be a \emph{MWC} graph, containing a set of configurations ($\mnode$) and edges ($\medge$).
Each configuration contains a data-structure ($\cn{content}$) and an amplitude field ($\cn{amp}$) having type $\rfun$, while each edge contains a weight field ($\cn{weight}$) having type $\rfun$.
All configurations in $\Phi$ have the same set $\dfun$ of data-points, where $\forall \phi\in\mnode.\cn{points}(\phi)=\dfun$ and $\cn{points}(\phi)\triangleq \{\cn{id}(\delta)|\delta \in \phi\}$.
\end{definition}

A MWC graph $\Phi$ essentially represents a quantum superposition state, while configurations are basis-ket states.
We assume that configurations and edges not mentioned in a MWC graph have zero amplitudes and weights, respectively.
In \Cref{fig:exmwc-intro}, we assume the case with all four nodes empty has amplitude $0$, and every connected edge has weight $0$ as well.

\subsection{Syntax and Typing}\label{sec:syntax}

\begin{figure}[t]
\[
\begin{array}{lrl}
\text{Metavariables} & & x, y \in \text{Variables},\quad
                         \zeta \in \cfun,\quad
                         m,n,j,k \in \mathbb{N},\quad
                         r \in \rfun,\quad
                         G \in \gfun(\delta) \\[4pt]
\text{Energy Data} &
v & ::= \; r \mid \cn{data}(\delta) \mid \cn{choice}(\delta)
\\[4pt]
\text{Energy expression}  & \varepsilon &::=\; v
  \mid \varepsilon + \varepsilon
  \mid \varepsilon \times \varepsilon
  \mid \cif{g}{\varepsilon}{\varepsilon} \\[4pt]
\text{Guard} &
  g &::=\; f(\varepsilon_1,\ldots,\varepsilon_k) \quad (f \in \Sigma)\\
  & &\phantom{::=}\; \mid\; \varepsilon = \varepsilon \; \mid\; g \wedge g \;\mid\; \neg g \\[4pt]

\text{Singleton constraint (U)} &
  \mu_U &::=\; \varepsilon \;\mid\; \partial x\in \dfun.\, \mu_U \\[4pt]
\text{Data constraint} &
  e_U &::=\; \mu_U \;\mid\; e_U + e_U \\[4pt]
\rowcolor{gray!15}
\text{Tuple} & \chi & ::= (x_1,\ldots,x_k) \\[4pt]
\rowcolor{gray!15}
\text{Constraint Flag} & \iota & ::= \; * \;\mid\; 0 \;\mid\; 1 \\[2pt]
\rowcolor{gray!15}
\text{Singleton constraint (S)} &
  \mu_S &::=\; \rset{\overline{x}}{\iota} \;\mid\; \partial \chi\in \dfun.\, \mu_S \\[2pt]
\rowcolor{gray!15}
\text{Relational constraint} &
  e_S &::=\; \mu_S \;\mid\; e_S + e_S \\[4pt]
\rowcolor{gray!15}
\text{Program} &
  P &::=\;\;
\{\;
\\\rowcolor{gray!15}
&& \qquad\quad \cn{struct}     : \gfun(\delta),\;\;
\\\rowcolor{gray!15}
&&  \qquad\quad \cn{choices}    : \cfun,\;\;
\\\rowcolor{gray!15}
&&  \qquad\quad \cn{initS}    : \Phi,\;\;
\\\rowcolor{gray!15}
&&  \qquad\quad \cn{constraint} : e_U,\;\;
\\\rowcolor{gray!15}
&&  \qquad\quad \cn{relation}   : e_S
\\\rowcolor{gray!15}
&& \qquad\}
\end{array}
\]
\caption{Grammar of \enqp.  Shaded productions are new relative
to \enq~\cite{10.1145/3747521}.  The original \enq program form
(without \textit{relation}) is a special case corresponding to
$e_S = \mathbf{Id}$, the trivial relational constraint whose
semantics is the uniform-superposition initial Hamiltonian. $\dfun$: a set of data-points.}
\label{fig:enqp-grammar}
\end{figure}

An \enqp program (\Cref{fig:enqp-grammar}) contains five components: a data-structure (\cn{struct}, $G$) containing a set of distinct data-points ($\delta$), a set of choices (\cn{choices}, $\cfun$), an initial state (\cn{initS}, $\Phi$) acting as the initial set of configurations with data-points being assigned with different choice values, a data constraint (\cn{constraint}, $e_U$), and a relational constraint (\cn{relation}, $e_S$).
The program syntax extends \enq with an initial starting state and a second syntactic category,
\emph{relational constraints}~$e_S$, that let the programmer
declare the structure of the initial Hamiltonian directly.
\Cref{fig:enqp-grammar} gives the extended grammar; shaded
productions are new.

When defining a data constraint ($e_U$), we assign an energy-loss value to unwanted cases.
Users are able to access the \cn{data} value of a data-point ($\cn{data}(\delta)$), turn a choice into an energy loss ($\cn{choice}(\delta)$), or use classical arithmetic ($\varepsilon$) and Boolean ($g$) expressions.
In \enqp, we allow a generic definition of energy loss, and the \enqp compiler finds the best fit by assigning real numbers to the generic energy loss, while ensuring every energy loss value is positive.

A \emph{relational constraint} $e_S$ is a sum of singleton relational constraints, allowing the programmer to specify relations among edges in MWC graphs.
Such a constraint $\rset{\overline{x}}{\iota}$ is defined by a list of variables ($\overline{x}$) and a relational flag $\iota$.
The former indicates the number of data points to examine together, and the latter indicates the movement relation among them; for example, \Cref{sec:constraintintro} shows how to examine every two data points with the swap relation between them.
The other two flags, $0$ and $1$, constrain whether the movement is in the same or a different direction, as in \Cref{sec:constraintintro}.

To specify the number of data points to examine, we permit binders, e.g., $\qbind{x}{\qbind{y}{\rset{x,y}{*}}}$.
In a \emph{singleton data constraint} $\mu_U$ and \emph{singleton relational constraint} $\mu_S$, users can bind a site tuple (a singleton tuple is a variable) via $\partial$-binders.
The binder $\qbinds{x}{\overline{x}}{\mu_S}$ or ($\qbinds{x}{\dfun}{\mu_U}$) binds a variable $x$ to a data-point, selected from the data-point set $\dfun$, and used in $\mu_S$ (or $\mu_U$).
In \enqp, we permit a choice to be a tuple data-structure, and users are able to mention a subcomponent of a tuple in a binder, e.g., $\qbinds{(x,y)}{\qbind{(u,v)}{\rset{y,v}{0}}}$ means that every choice is a pair and we establish the relation on the second part of each pair as $0$-flagged.

The $\partial$-binders quantify the atom over all tuples of nodes defined in $\overline{x}$, a data-point subset of $\cn{struct}$, producing a sum of
two-site operators that define the full initial Hamiltonian.
One can omit the set notation in a binder ($\qbind{\chi}{\mu}$) to refer to choosing $x$ from the whole data-point set, and $\qbind{\chi}{\qbind{\chi'}{\mu}}$ selects two tuples $\chi$ and $\chi'$, where they must be distinct data-points.

\myparagraph{Typing}\label{sec:typing}
\Cref{fig:exp-proofsystem-1} defines the type system, which has the top-level judgment $\vdash P$, as well as the expression judgment $\Gamma\vdash -:\tau$, where $\Gamma$ is the type environment, mapping from variables to data-point sets; $-$ could be a program $P$, an arithmetic ($\varepsilon$) and Boolean ($g$) expression; $\tau$ is the type construct, defined as follows.

{\small
\[\tau := \rfun \mid \cn{bool}\mid \sfun
\]
}
Here, $\rfun$, $\cn{bool}$, and $\sfun$ indicate an arithmetic expression, a Boolean expression, and a constraint expression, respectively.

\begin{figure*}[t]
{\small
  \begin{mathpar}
            \inferrule[T-Data]{}{x:D,\Gamma \vdash \cn{data}(x):\rfun}
            \qquad
              \inferrule[T-Choice]{}{x:D,\Gamma \vdash \cn{choice}(x):\rfun}
                \qquad
                 \inferrule[T-Plus]{\Gamma \vdash \varepsilon_1 : \rfun\quad\Gamma \vdash \varepsilon_2 : \rfun}{\Gamma \vdash \varepsilon_1 + \varepsilon_2:\rfun}
\qquad
                 \inferrule[T-Times]{\Gamma \vdash \varepsilon_1 : \rfun\quad\Gamma \vdash \varepsilon_2 : \rfun}{\Gamma \vdash \varepsilon_1 \times \varepsilon_2:\rfun}

             \inferrule[T-If]{\tjudge{\Gamma}{}{g}{\cn{bool}}\\ \tjudge{\Gamma}{}{\varepsilon_1}{\rfun}\\\tjudge{\Gamma}{}{\varepsilon_2}{\rfun}}{\Gamma \vdash \ife{g}{\varepsilon_1}{\varepsilon_2}:\rfun}

        \inferrule[T-App]{\tjudge{\Gamma}{}{\varepsilon_1}{\rfun}\;\;\ldots\;\;\tjudge{\Gamma}{}{\varepsilon_k}{\rfun}}{\Gamma \vdash f(\varepsilon_1,\ldots,\varepsilon_k):\cn{bool}}

   \inferrule[T-Eq]{\Gamma \vdash \varepsilon_1 : \rfun\\\Gamma \vdash \varepsilon_2 : \rfun}{\Gamma \vdash \varepsilon_1 = \varepsilon_2:\cn{bool}}

     \inferrule[T-And]{\Gamma \vdash g_1 : \cn{bool}\\\Gamma \vdash g_2 : \cn{bool}}{\Gamma \vdash g_1 \wedge g_2:\cn{bool}}

\inferrule[T-Neg]{\Gamma \vdash g : \cn{bool}}{\Gamma \vdash \neg g:\cn{bool}}

          \inferrule[T-Rel]{ }{\tjudge{\overline{x}:\overline{\dfun}}{}{\rset{\overline{x}}{\iota}}{\sfun}}

                     \inferrule[T-Sum]{\emptyset \vdash e_1 : \sfun\\\emptyset \vdash e_2 : \sfun}{\emptyset \vdash e_1 + e_2:\sfun}

 \inferrule[T-Bind]{\forall \dfun_a\in\overline{\dfun_1}.\,\slen{\overline{x}}<\slen{\dfun_a}\\ \forall i.\overline{\dfun}_i \subseteq \overline{\dfun}_{i+1}\\\forall \dfun_a\in\overline{\dfun_1}.\,\dfun_a\subseteq \overline{\dfun}_0 \\  \overline{\dfun}_{\slen{\overline{\dfun}}-1}\cap \cn{range}(\Gamma)=\emptyset \\\tjudge{\overline{y}:\overline{\dfun_1},\overline{x}:\overline{\dfun},\Gamma}{}{e}{\sfun} }{\tjudge{\overline{x}:\overline{\dfun},\Gamma}{}{\qbind{(\overline{y})\in (\overline{\dfun_1})}{e}}{\sfun}}

    \inferrule[T-Top]{\dom{e_U}\cup \dom{e_S} \vdash \Phi\\\tjudge{\emptyset}{}{e_U}{\sfun}\\\tjudge{\emptyset}{}{e_S}{\sfun}}
                {\vdash \{\cn{struct}:\gfun(\delta),\cn{choices}:\cfun,\cn{initS}:\Phi,\cn{constraint}:e_U,\cn{relation}:e_S\}}
  \end{mathpar}
}
\caption{Type rules.}
\label{fig:exp-proofsystem-1}
\end{figure*}

The type system guarantees three properties. First, it guarantees the right types for the right expressions: rule \rulelab{T-Plus} types the addition expression with the $\rfun$ type, rule \rulelab{T-And} types a Boolean expression, and rule \rulelab{T-Sum} types a constraint choice operator.
Second, the type system ensures that type variables are properly scoped.
Rules \rulelab{T-Data} and \rulelab{T-Choice} ensure variables are properly scoped in the type environment, and rule \rulelab{T-Bind} binds a variable in the type environment and maps it to its right data-point set.
Rules \rulelab{T-Sum} and \rulelab{T-Top} ensure that the top-level constraints have no free variables.

Third, we ensure that constraint variables are disjoint and that no constraints are empty. In \enqp, nested constraint variables must represent distinct data points; e.g., $\qbind{x}{\qbind{y}{\mu}}$ means $x$ and $y$ must be two distinct data points.
Rule \rulelab{T-Rel} partitions the type environment ($\overline{x}:\overline{\dfun}$) into distinct subsets to type the subterms.

In rule \rulelab{T-Bind}, we use several premises to compute the correct data-point set for the variable $y$, because selecting the data-point set might empty a constraint.
For example, since $\qbinds{x}{\dfun}{\qbinds{y}{\dfun}{\mu}}$ requires that $x$ and $y$ to be two distinct data-points, if $\slen{\dfun} = 1$, the constraint becomes empty.
To avoid an empty constraint, we ensure that all data points mentioned in the type environment can be partitioned into regions, and that $y$'s data-point set $\dfun$ lies within a region $\overline{\dfun}$ with variable set $\overline{x}$. We also ensure that the sets $\overline{\dfun}$ can be ordered as $\overline{\dfun}_0\subseteq ...\subseteq\overline{\dfun}_{\slen{\overline{\dfun}}-1}$, and every set in $D=\{\overline{\dfun_1}\}$ is a subset of $\overline{\dfun}_0$. Then, the cardinality of $\overline{\dfun_1}$ exceeds the number of variables ($\overline{x}$ and $\overline{y}$), ensuring no empty constraint.
In rule \rulelab{T-Top}, $\dom{e_U}\cup \dom{e_S}\vdash \Phi$ collects all the mentioned data points and ensures that the configurations in $\Phi$ cover the data points.

We discuss type soundness in the next section.

\subsection{Semantics}\label{sec:sema}

The semantics of \enqp is divided into two categories: constraint semantics and dynamic semantics, as discussed below.

\myparagraph{Constraint Semantics.}\label{sec:constrainsem}
Constraint semantics is a function that yields the judgment $\denote{P}^{\Phi} = \Phi'$, where $P$ is a program and $\Phi$/$\Phi'$ are MWC graphs.
A data constraint might contain an expression $\varepsilon$ with the type $\ufun(\overline{x})$, where we can interpret $\overline{x}$ with $\overline{\delta}$, so $\varepsilon[\overline{\delta}/\overline{x}]$ produces an energy loss $r$; such rules are shown in \Cref{fig:equiv}.

We now define the constraint semantics as a function interpreting a MWC graph $\Phi=(\mnode, \medge)$, with each node $\phi\in\mnode$ having a fixed set of data-points $\overline{\delta}$, coming from the user-defined data structure $\cn{struct}(P)$.
For simplicity, we linearize the tuples in each binding, as in the following example.
{
\[\qbinds{(\overline{x})}{(\overline{D_1})}{\qbinds{(\overline{y})}{(\overline{D_2})}{e}}=\qbinds{(\overline{x},\overline{y})}{(\overline{D_1},\overline{D_2})}{e}\]
}

We first define the data constraint semantics.

\begin{figure}[t]
\[\small
\begin{array}{l}
\denote{r}^{v}= r
\quad
\denote{\cn{data}(\delta)}^{v}=\cn{data}(v(\delta))
\quad
\denote{\cn{choice}(\delta)}^{v}=\cn{choice}(v(\delta))
\quad
\denote{\varepsilon_1 + \varepsilon_2}^{v}=\denote{\varepsilon_1}^{v} + \denote{\varepsilon_2}^{v}
\\[0.5em]
\denote{\varepsilon_1 \times \varepsilon_2}^{v}=\denote{\varepsilon_1}^{v} \times \denote{\varepsilon_2}^{v}
\quad
\denote{\ife{g}{\varepsilon_1}{\varepsilon_2}}^{v} = \ife{\denote{g}^{v}}{\denote{\varepsilon_1}^{v}}{\denote{\varepsilon_2}^{v}}
\quad
\denote{\neg \varepsilon}^{v} = \neg \denote{\varepsilon}
\\[0.5em]
\denote{f(\varepsilon_1,\ldots,\varepsilon_k)}^{v}=f(\denote{\varepsilon_1}^{v},\ldots,\denote{\varepsilon_k}^{v})
\quad
\denote{\varepsilon_1 = \varepsilon_2}^{v} = \denote{\varepsilon_1}^{v} = \denote{\varepsilon_2}^{v}
\quad
\denote{\varepsilon_1 \wedge \varepsilon_2}^{v} = \denote{\varepsilon_1}^{v} \wedge \denote{\varepsilon_2}^{v}
\end{array}
\]
\caption{Expression Level Constraint Semantics}
\label{fig:equiv}
\end{figure}

\begin{definition}[Data Constraint Semantics]\label{def:datasem}\rm
Given a data constraint $e_u$ and a MWC graph $\Phi=(\mnode, \medge)$, we define data-constraint interpretation function $\denote{-}^{\phi}$, adding the amplitude ($\cn{amp}$) for every $\phi\in \mnode$, as $\mnode[\phi \mapsto \phi[\cn{amp}(\phi)\mapsto\denote{e_U}^{\phi}+\cn{amp}(\phi)]]$, where we define the rules as follows.

{
       \begin{mathpar}
        \inferrule[]{}{ \denote{e_1 + e_2}^{\phi} := \denote{e_1}^{\phi} \,\textcolor{purple}{+}\, \denote{e_2}^{\phi} }

        \inferrule[]{}
        { \denote{\qbinds{\overline{x}}{\overline{\dfun}}{\varepsilon}}^{\phi} := \textcolor{purple}{\Msum}_{\overline{\delta}\in \cn{gen}(\overline{\dfun},\cn{points}(\phi))} \denote{\varepsilon[\overline{\delta}/\overline{x}]}^{\phi} }
       \end{mathpar}
}

Here, $\denote{\varepsilon}^{\phi}$ is the interpretation function in \Cref{fig:equiv}, $\textcolor{purple}{+}$ is a function sum, with $f_1+f_2 := \forall \phi\in \mnode. \mnode[\cn{amp}(\phi)\mapsto f_1(\phi)+f_2(\phi)]$, and the  $\cn{gen}$ function is defined as follows.

{\small
\[
\cn{gen}(\overline{\dfun}_0,\dfun'_0) := \overline{\dfun}_0\cap D'_0
\wedge
\forall j.\, \cn{gen}(\overline{\dfun}_{j+1},\dfun'_{j+1}) := \overline{\dfun}_{j+1}\cap (\dfun'_j\textbackslash \{\delta_j\})
\]
}
\end{definition}

The iterative function $\cn{gen}$ generates a list of data points ($\overline{\delta}$) for calculating the energy loss values.
Every singleton data constraint has the form $\qbinds{x_0}{\dfun_0}{...\qbinds{x_n}{\dfun_n}{\varepsilon}}$, and each variable $x_j$ is selected from the set $\dfun_j$ subtracting the data-points selected for $x_0,...,x_{j-1}$.

We define the relational constraint below as the production of the edge set $\medge$, i.e., configuration pairs connected by edges.

\begin{definition}[Relational Constraint Semantics]\label{def:datasem}\rm
Given a constant weight $r_a$, given a data constraint $e_s$ and a MWC graph $\Phi=(\mnode, \medge)$, we define data-constraint interpretation function $\denote{-}^{(\phi,\phi')}$, adding the weight ($\cn{weight}$) for every edge $(\phi,r,\phi')\in \medge$, as $\medge[(\phi,r,\phi') \mapsto (\phi,r+r_a,\phi)]$ if $\denote{e_s}^{(\phi,\phi')}$ is \cn{true}, where we define the rules as follows.

{\small
\[
\begin{array}{r@{\;}c@{\;}ll}
         \denote{e_1 + e_2}^{(\phi,\phi')} &:=& \denote{e_1}^{(\phi,\phi')} \,\textcolor{purple}{+}\, \denote{e_2}^{(\phi,\phi')}
                  \\[0.5em]
 \denote{\qbinds{\overline{x}}{\overline{\dfun}}{\mu}}^{(\phi,\phi')} &:=& \textcolor{purple}{\sum}_{\overline{\delta}\in \cn{gen}(\overline{\dfun},\cn{points}(\phi))} \denote{\mu}^{(\overline{\delta_1},\overline{\delta_2})} &
          \cn{when}\,\,\cn{points}(\phi)=\overline{\delta_1}\cup \overline{\delta'}
\\[0.2em]
&&&\qquad\qquad\wedge \cn{points}(\phi')=\overline{\delta_2}\cup\overline{\delta'}
         \wedge \cn{id}(\overline{\delta_1}) = \cn{id}(\overline{\delta_2})
        \end{array}
\]
}

Here, $\textcolor{purple}{+}$ is a function sum, with $f_1+f_2 := \forall \phi\in \mnode. \medge[(\phi,r,\phi') \mapsto (\phi,f_1(r)+f_2(r),\phi)]$, and the Boolean function $\denote{\mu}^{(\overline{\delta_1},\overline{\delta_2})}$ is defined as follows.

{\small
\[
\begin{array}{l@{\;}c@{\;}l}
 \inferrule[]{}{ \denote{\rset{\overline{x}}{0}}^{(\overline{\delta_1},\overline{\delta_2})}} &:=&
 \exists n.\,\forall i.\,\cn{choice}(\delta_1[i]) = \cn{choice}(\delta_2[i])  + n
 \\[0.8em]
  \inferrule[]{}{ \denote{\rset{\overline{x}}{1}}^{(\overline{\delta_1},\overline{\delta_2})}} &:=&
 \forall n.\,\forall i.\,\cn{choice}(\delta_1[i]) \neq \cn{choice}(\delta_2[i])  + n
 \\[0.8em]
  \inferrule[]{}{ \denote{\rset{\overline{x}}{*}}^{(\overline{\delta_1},\overline{\delta_2})}} &:=&
   \cn{perm}^{\slen{\overline{x}}}(\overline{\delta_1},\overline{\delta_2})
\end{array}
\]
}
\end{definition}

The function $\cn{perm}^{n}(\overline{\delta_1},\overline{\delta_2})$ checks whether $\overline{\delta_1}$ and $\overline{\delta_2}$ are an $n$-order permutation, which refers to that $\overline{\delta_1}$ and $\overline{\delta_2}$ have the same set of choices, and there are $n$ positions in $\overline{\delta_1}$ and $\overline{\delta_2}$ having different choices. For example, $1,2,3$ and $1,3,2$ is a $2$-order permutation while $1,2,3$ and $2,3,1$ is a $3$-order permutation.

Finally, the constraint semantics is a function $\denote{P}^{(\mnode,\medge)}$, defined as follows.

{
\[
\denote{e_U}^{(\mnode,\medge)} := \textcolor{purple}{\sum}_{\phi\in \mnode}\denote{e_U}^{\phi}
\qquad
\denote{e_S}^{(\mnode,\medge)} := \textcolor{purple}{\sum}_{(\phi,r,\phi')\in \medge}\denote{e_S}^{(\phi,\phi')}
\]
}

With the constraint semantics, we show the type soundness theorem below.

\begin{theorem}[Type Soundness]\label{thm:typesound}\rm
For a program $P$, with $\vdash P$, where $\Phi=(\mnode,\medge)$ is the initial state, there are valid interpretations of $\denote{P}^{\Phi}$.
\end{theorem}

\begin{proof}
We prove by induction on the type rules in \Cref{fig:exp-proofsystem-1}.
\end{proof}

\myparagraph{Dynamic Semantics.}
The constraint semantics of an \enqp program ($P$) is a function $\denote{-}^{\Phi}$ on the MWC graph $\Phi = (\mnode,\medge)$, and the dynamic semantics applies the AQC evolution on the graph.
The result follows the AQC theorem: given an initial ground state (minimized eigenstate) of the initial Hamiltonian, i.e., the relational constraint, AQC evolution slowly drives the state toward the ground state of the final Hamiltonian, i.e., the data constraint.
In \enqp, users can define initial states, while the AQC theorem guarantee holds only for the initial state that is the ground state of the relational constraint.

The AQC theorem unveils the result of applying an \enqp program in an AQC system.
However, it does not unveil the dynamic semantics of a single-step evolution.
Essentially, the AQC evolution follows $\hat{H}(t)=(1-t)\hat{H}_i + t\hat{H}_f$, where $\hat{H}_i$ comes from the relational constraint and $\hat{H}_f$ from the data constraint.

In \enqp, the dynamic semantics are theorems derived from the constraint semantics and the AQC evolution theorem. We list them as theorems below.
To examine a single transition step, we first examine the evolution of individual constraints, yielding two rotation functions, $\psi$ and $\theta$, that produce rotations in $[0,\frac{\pi}{2})$.
Here, we mainly develop the theory for the system constructed by \enqp and evolved by Hamiltonian simulation ($e^{i\hat{H} t}$), though other AQC systems may apply.
Given a program $P$ and an MWC graph $\Phi=(\mnode,\medge)$, the definition partitions configurations and edges into relevant and irrelevant sets.
Given a data constraint ($e_U$), the relevant sets contain configurations that yield a non-zero value when $e_U$ is applied, while configurations in the irrelevant set yield a $0$ when $e_U$ is applied.

\begin{theorem}[Data Constraint Evolution]\label{thm:dataevolve}\rm
Given a data constraint $e_U$ and an $n$-dimensional MWC graph $\Phi=(\mnode,\medge)$, for every $\phi\in \mnode$, the instantaneous effect $\phi'$, evolving the system for an evolution time $s$, compilation factor $\omega$, on the relevant set is an update on $\phi$'s amplitude, as $\cn{amp}(\phi') = e^{i\psi(s)}*\cn{amp}(\phi)$, where $\psi(s) \propto\frac{s* \omega (\cn{amp}(\denote{e_U}^{\phi})-\cn{amp}(\phi))}{n}$. The effect on the irrelevant set is neutral, as $\cn{amp}(\phi') = \cn{amp}(\phi)$.
\end{theorem}

The relevant sets have a different definition for a relational constraint ($e_S$). Essentially, $e_S$ is a Boolean function determining whether an edge, containing two configurations, is in the relevant set or in the irrelevant set.

\begin{theorem}[Relational Constraint Evolution]\label{thm:relevolve}\rm
Given a relational constraint $e_S$ and an $n$-dimensional MWC graph $\Phi=(\mnode,\medge)$, for every $(\phi,r,\phi')\in \medge$, the instantaneous effect $r'$, evolving the system for a small time $s$, compilation factor $\omega$, on the relevant set is as $r' = e^{i\theta(s)}*r$, where $\theta(s) \propto\frac{s * \omega * m}{n^2}$, where $m$ is the number of nonzero increment in $\medge$, i.e., the number of edges $(\phi,r,\phi')$ such that $\cn{weight}(\denote{e_S}^{(\phi,\phi')})-r\neq 0$. The effect on the irrelevant set is neutral, as $r' = r$.
\end{theorem}

The above unveils the dynamic semantics for evolving the data and relational constraints individually, yielding two angle functions $\psi(s)$ and $\theta(s)$ in the range $[0,\frac{\pi}{2})$, respectively interpreting each configuration and edge in a MWC graph. In fact, both angles must be $\ll \frac{\pi}{2}$.

With the definition for individual constraints, we define the dynamic semantics of the whole system, based on two-step operations, $\cn{step}^{b}(\psi(s),\theta(s))$ and $\cn{step}^{g}(\psi(s),\theta(s))$, which are two monotonic functions within the AQC evolution steps, i.e., given a single step time factor $\Delta t$, each evolution stops exactly at the $\lceil\frac{1}{\Delta t}\rceil$ steps.
The former defines the step behavior of a relevant configuration, essentially a bad state with energy loss, and the latter defines the step behavior of a good state, an irrelevant configuration with $0$ energy loss.
The step evolution of the two operations modifies the likelihood of good and bad states.
In each step, we modify $\psi$'s time factor from $0$ to $1$ and $\theta$'s factor from $1$ to $0$ by $\Delta t$.

\begin{theorem}[Dynamic Semantics]\label{def:dynsem}\rm
Given an initial state as MWC graph $\Phi=(\mnode,\medge)$, for every $\phi\in\mnode$, we have an angle $\psi(s)$, and for every relevant edge $(\phi,r,\phi')$, we have an angle function $\theta(s)$; we define the dynamic transition steps based on relevant edges, $\cn{step}^{b}(\psi(s),\theta(s))$ and $\cn{step}^{g}(\psi(s),\theta(s))$, as the likelihood multiplication on configuration's amplitude, i.e., given $n$ step, the evolution transitions $\phi$'s amplitude as $\cn{step}^{g}_{n}*\cn{amp}(\phi)$ for good states and the amplitude $\cn{step}^{b}_{n}*\cn{amp}(\phi)$ for bad states.

{
\[
\begin{array}{l}
\cn{step}^{b}_{0} = 1
\qquad
\cn{step}^{g}_{0} = 1
\\[0.5em]
\cn{step}^{b}_{j+1} = e^{i\psi(j\Delta t)} \cos{[\theta(1-j\Delta t)]} * \cn{step}^{b}_j - i \sin{[\theta(1-j\Delta t)]} * \cn{step}^{g}_j
\\[0.5em]
\cn{step}^{g}_{j+1} = \cos{[\theta(1-j\Delta t)]} * \cn{step}^{g}_j - e^{i\psi(j\Delta t)} i \sin{[\theta(1-j\Delta t)]} * \cn{step}^{b}_j
\end{array}
\]
}
\end{theorem}

\begin{proof}
The proofs of \Cref{thm:dataevolve,thm:relevolve,def:dynsem} rely on the case analysis on individual configurations and edges.
We assume a restricted AQC where the program data constraints are diagonal and the compilation is Hamiltonian-simulation-based; therefore, we can analyze \Cref{thm:dataevolve,thm:relevolve} by applying Hamiltonian-simulation operators to compiled Pauli-string programs. The proof for \Cref{def:dynsem} alters the effects of the two simulation results in \Cref{thm:dataevolve,thm:relevolve}.
\end{proof}

Based on \Cref{def:dynsem}, we have the following lemma regarding the monotonicity of the two-step operations.

\begin{theorem}[Step Operation Monotonicity]\label{thm:max}\rm
For the evolution steps $\lceil\frac{1}{\Delta t}\rceil$, i.e., $j\in[0,\lceil\frac{1}{\Delta t}\rceil)$, $\cn{step}^{b}_{j}$ is a monotonic decreasing function, while $\cn{step}^{g}_{j}$ is a monotonic increasing function.
\end{theorem}

\begin{proof}
The proof proceeds by induction on the steps for $\cn{step}^{b}$ and $\cn{step}^{g}$.
\end{proof}

\Cref{def:dynsem,thm:max} cover the case where all edges are relevant.
They also assume that relevant and irrelevant edges are orthonormal. As shown in \Cref{sec:example}, the \enqp relational constraints are compiled to $\uparrow$ and $\downarrow$, which might modify basis-vectors, i.e., an edge in the relevant set might be modified to one in the irrelevant set.
In this paper, we assume such a case does not occur and leave it as future work.

To ensure that all nonzero weighted edges are relevant, the initial MWC graph is the eigenstate of the relational constraint.
An interesting study is what if we do not start with the right initial state.
Some initial state preparations, such as states of distinct elements, are costly to implement on a quantum computer.
We show in the good approximation theorem below that starting with an ideal relational constraint but an easily prepared, imperfect initial state—one that is not the eigenstate of the relational constraint—can still produce meaningful results.

Assume that $\Phi_1=(\mnode_1,\medge_1)$ is the eigenstate of a relational constraint ($e_S$), where $\mnode_1 \subseteq \mnode$.
We define such a relation as $\Phi_1\sqsubseteq_{e_S}\Phi$, with $\Phi_1$ being the eigenstate of $e_S$ and all non-zero configurations and edges in $\Phi_1$ are defined in $\Phi$.
Under this definition, the following theorem relates $\Phi_1$ and $\Phi$.

\begin{theorem}[Good Approximation]\label{thm:aqcorder1}\rm
Given $P$, with $e_S$ being the relational constraint and $\Phi_1=(\mnode_1,\medge_1)$ being the initial state, assume $\Phi=(\mnode,\medge)$, with $\Phi_1\sqsubseteq_{e_S}\Phi$, $P$ partitions $\mnode$ into relevant ($\mnode_r$) and irrelevant ($\mnode_i$) sets, as $P$ keeps the two sets being orthonormal; for every $\phi \in \mnode_i$, the AQC evolution results keep the final amplitudes the same as the initial one in $\Phi$.
\end{theorem}

\begin{proof}
The proof follows through the dynamic semantics analysis in \Cref{thm:dataevolve,thm:relevolve,def:dynsem}.
The key is that Hamiltonian-simulation-based AQC requires alternating between the two constraints to drive state evolution. If a configuration (or configurations connected by an edge) is only relevant for a single constraint (data or relational), its AQC evolution effect is only a phase rotation.
\end{proof}

\section{The \enqp Compilation Procedure}\label{sec:compilation}

\begin{figure}[t]
    \centering
    \includegraphics[width=1\textwidth]{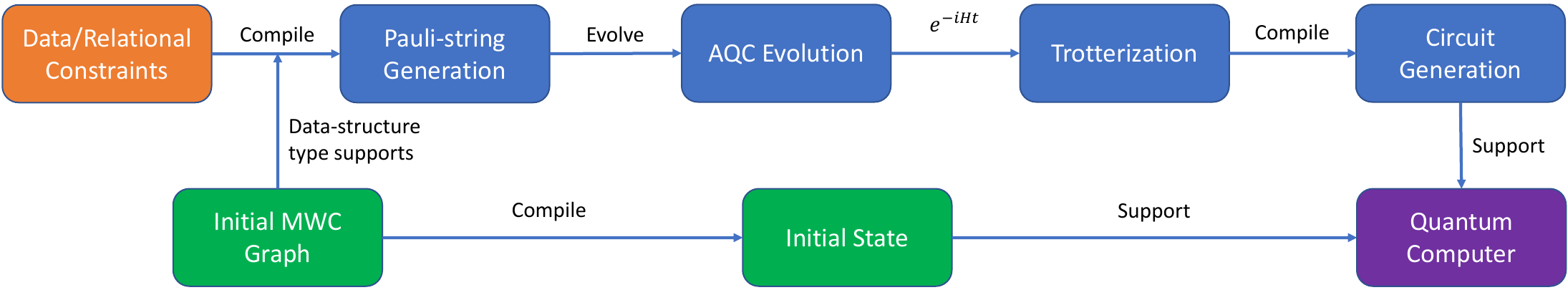}
    \caption{The ENQ+ Compiler Flow}
    \label{fig:enqcompiler}
\end{figure}

This section introduces the \enqp compilation procedure, shown in \Cref{fig:enqcompiler}, which can be divided into the following steps.

\begin{itemize}
\item We compile the initial MWC graph into an initial quantum state, prepared as a superposition containing all candidate solutions, as the eigenstate of the relational constraint.
\item Translate both the relational and data constraints, written in the syntax in \Cref{fig:enqp-grammar}, to Pauli-string-based Hamiltonians.
\item Guide the system toward that ground state, either via gate-based AQC, compiled by Trotterization.
\end{itemize}

\subsection{Generating Initial Quantum State}\label{sec:initialstate}

The classical implementation explicitly traverses all possible choices to build the solution space. For instance, a list of $n$ variables with two options each would generate $N = 2^n$ combinations.
Instead of iterating through each combination, we prepare a quantum superposition that represents all $N$ possible solutions simultaneously.
We allocate a set of qubits to encode the choice variables (e.g., one qubit per binary choice).

In AQC, we start with an initial quantum superposition state $\ket{\varphi(0)}$, the eigenstate of the relational constraint.
There is no algorithm to prepare an arbitrary quantum state, but there are some algorithms to prepare special kinds.
In \enqp, we allow users to start with several initial states, including the uniform superposition state, Hamming weight state, permutation state, and distinct element state.
For each state, we can attach special amplitudes to different basis-ket states.

{
\begin{center}
$
\ket{\varphi}=\sum_{j=0}^{2^n\sminus 1}\alpha_j\ket{b_j}
\qquad
\qquad
\ket{\varphi(0)}=\sum_{j=0}^{2^n\sminus 1}\frac{1}{\sqrt{2^n}}\ket{j}
$
\end{center}
}

We provide two examples here.
A typical initial quantum state is a uniform superposition. For example, if a variable has two choices, a single qubit in state $\ket{0}$ is transformed to $\frac{1}{\sqrt{2}}(\ket{0} + \ket{1})$. For $n$ binary variables, this yields a uniform superposition of all $2^n$ combinations, which can be transformed to the form listed on the right side above. The state has $2^n$ different basis-kets, each with amplitude being $\frac{1}{\sqrt{2^n}}$ and a basis-vector $\ket{j}$.
Here, $j$ denotes the binary string representation of the number $j$.
In the quantum setting of the clique-finding example, we want to start with an $n$-qubit initial superposition state of $K$-th Dicke state, $\begin{pmatrix} n \\ K \end{pmatrix}^{-\frac{1}{2}}\sum_{h(j)=K}\ket{j}$, where $h(j)$ checks the bitstring representation of $j$ having exactly $K$ copies of $1$.
Different selections of initial states might affect the convergence speed of the AQC computation.

\subsection{Compiling EnQ+ Programs to Pauli-strings}\label{sec:initham}

\begin{figure*}[t]
{\small
  \begin{mathpar}
      \inferrule[C-EU]{FV(\varepsilon)=\overline{x}}
            {\tcompile{\overline{x}:\overline{\dfun},\Gamma}{}{\varepsilon}{\rfun}{\Msum_{\overline{k}\in\overline{C}}-\cn{fit}(\varepsilon[\overline{k}/\overline{x}])\cdot \bigcirc_{x\in \overline{x}} \denote{\ket{k}}^C(x)} }

                  \inferrule[C-Bind]{\tcompile{y:D,\overline{x}:\overline{D},\Gamma}{}{e}{\sfun}{\hat{H}}}
            {\tcompile{\overline{x}:\overline{\dfun},\Gamma}{}{\qbind{y\in \dfun}{e}}{\sfun}{\Msum_{y\in \dfun\textbackslash \{\overline{x}\}}\hat{H}(y)} }

          \inferrule[C-ES]{ }
            {\tcompile{\overline{x}:\overline{\dfun},\Gamma}{}{\rset{\overline{x}}{\iota}}{\sfun}{ \Msum_{\overline{\delta}\in\cn{gen}(\overline{\dfun},\cn{points}(\cn{struct}(P)))}\denote{\rset{\overline{\delta}}{\iota}}^C} }

      \inferrule[C-Sum]{\tcompile{\emptyset}{}{e_1}{\sfun}{\hat{H}_1}\\\tcompile{\emptyset}{}{e_2}{\sfun}{\hat{H}_2}}
            {\tcompile{\emptyset}{}{e_1+e_2}{\sfun}{\hat{H}_1 + \hat{H}_2}}
  \end{mathpar}
}
\vspace*{-0.3em}
\caption{Compilation Rules. $P$ is the program and $\cn{points}$ produces data-points of a configuration in $\cn{struct}(P)$.}
\label{fig:exp-transform}
\vspace*{-0.8em}
\end{figure*}

We examine the compilation of constraint expressions $e_U$ and $e_S$ to Pauli-string forms $\hat{H}$, as the extension of the \enqp type system, $\tcompile{\Gamma}{}{e}{\sfun}{\hat{H}}$, with $\hat{H}$ having the form $\sum_j \otimes_{k=0}^n \pau_k$. The rules are in \Cref{fig:exp-transform}.

Rule \rulelab{C-Sum} turns an \enqp composition operation into a linear sum; \rulelab{C-Bind} turns the data-point choice for $y$ into a linear sum over $y$'s possible selections, while \rulelab{C-EU} and \rulelab{C-ES} are respectively the data and relational constraint rules for a fixed selection of data-points referred to by $\overline{x}$.
We discuss the data and constraint individual compilation functions below.

\myparagraph{Marking Choice States.}
The key step in generating Pauli strings turns basis-vector states into Pauli-string operations.
As demonstrated in \Cref{sec:gap}, given an input/output pair $\ket{b} \to \ket{b'}$ ($b\in[0,1]$), we examine the $(b,b')$ pair and yield an operator in $\{\uparrow,\downarrow,\zeros,\ones\}$. We name the procedure state marking, because applying an operator to an unqualified vector yields $\zero$, e.g., $\zeros\ket{1} = \zero$.

The procedure extends to the high-order case.
In \enqp, a solution for a problem contains many nodes, each of which needs some qubits to represent a choice.
Given a choice set $C$, we use a $\ulcorner \log(\slen{C}) \urcorner$ number of qubits to represent all the choices.
For example, in the graph coloring problem, users might want to use a choice set of colors, $C=\{\cn{Black},\cn{Red}, \cn{Blue}, \cn{White}\}$. Here, we can use two qubits per node to represent the four colors as $\ket{00}$, $\ket{01}$, $\ket{10}$, and $\ket{11}$, respectively.
We can extend this idea to an $n$-qubit state marking, $\ket{j}_n \to \ket{k}_n$, using the following operations.

{\small
\begin{equation}
\begin{array}{l@{\;}c@{\;}l}
\denote{\ket{i} \to \ket{j}}^{C} &\triangleq& \bigotimes_{u=0}^{\ulcorner \log(\slen{C}) \urcorner\sminus 1}\mathpzc{A}^u(i,j)
\end{array}
\label{eq:makingstate}
\end{equation}
}
where
{\small
\[
\begin{array}{c}
\textcolor{spec}{\ket{j}_n} = \textcolor{spec}{\ket{\tob{j}[0]}\ket{\tob{j}[1]}...\ket{\tob{j}[n\sminus 1]}}
\\[0.2em]
\begin{array}{l@{\;}c@{\;}l@{\;\;}c@{\;}l@{\qquad\quad}l@{\;}c@{\;}l@{\;\;}c@{\;}l}
\textcolor{spec}{\mathpzc{A}^k(j,m)}&=&\textcolor{spec}{\uparrow}&\cn{if}&\textcolor{spec}{\tob{j}[k]=0 \wedge \tob{m}[k]=1}
&
\textcolor{spec}{\mathpzc{A}^k(j,m)}&=&\textcolor{spec}{\mathbb{1}}&\cn{if}&\textcolor{spec}{\tob{j}[k]=\tob{m}[k]=1}
\\[0.2em]
\textcolor{spec}{\mathpzc{A}^k(j,m)}&=&\textcolor{spec}{\downarrow} & \cn{if} & \textcolor{spec}{\tob{j}[k]=1 \wedge \tob{m}[k]=0}
&
\textcolor{spec}{\mathpzc{A}^k(j,m)}&=&\textcolor{spec}{\mathbb{0}}&\cn{if}&\textcolor{spec}{\tob{j}[k]=\tob{m}[k]=0}
\end{array}
\end{array}
\]
}

Here, $\tob{j}$ converts $j$ to a bitstring, and the operations in $\{\uparrow,\downarrow,\zeros,\ones\}$ can be compiled to Pauli operators as follows.

{\small
\[
\uparrow\, \triangleq X-iY
\qquad
\downarrow\, \triangleq X+iY
\qquad
\zeros \triangleq I + Z
\qquad
\ones \triangleq I - Z
\]
}

The state marking function ($\denote{\ket{i} \to \ket{j}}^{C}$) essentially utilizes the bitstring representations of $i$ and $j$.
For the $k$-th bit position ($\tob{i}[k]$ and $\tob{j}[k]$), we compare the values and produce a Pauli operator for the position.

\myparagraph{Compiling data constraints.}
The compilation of data constraints directly utilizes \Cref{eq:makingstate}.
Given a choice set $C$, to mark a specific state as $k\in C$, we model $\denote{\ket{k}}^C(x)$ in rule \rulelab{C-EU} as follows.

\[\denote{j}^C(x) \triangleq \denote{\ket{k} \to \ket{k}}^{C}(x)\]

Here, we include the data-point location ($x$) as the target for applying the generated Pauli strings in $\denote{\ket{k} \to \ket{k}}^{C}$.
The end result of compiling a data constraint is a final Hamiltonian $H_f$, as a sum of weighted Pauli operators (single-qubit $Z$s, two-qubit $Z\otimes Z$ interactions, etc., depending on the structure of the data constraints).

The above generate Pauli strings representing non-zero data points, with $\cn{fit}(\varepsilon[\overline{k}/\overline{x}])$ assigning the energy scores for these data points.
The \cn{fit} function interprets the expression $\varepsilon$ by assigning variables to numbers representing data-point locations.
Additionally, for generic energy-score constructs ($\overline{v}$) in $\varepsilon$, we try different real-number assignments ($\overline{r}$) and minimize the steps using the $\cn{step}^{g}$ function for estimation. We list the minimization function below.

{
\[
\min_{n,\overline{r}}\cn{step}^{g}_n(\psi[\overline{r}/\overline{v}](s),\theta(s))
\]
}

Finally, the negative sign in \rulelab{C-EU} ensures the correctness for the Hamiltonian-simulation-based AQC compilation.

\myparagraph{Compiling relational constraints.}
We mainly explain $ \denote{\rset{\overline{\delta}}{\iota}}^C$ here.
We first define useful helper functions.
For the vector $\neq i$, the function rewrites those cases to $\zero$.
For an arbitrary input, we can utilize linear sum operations to construct the following operator.

{\small
\[
\denote{+k}^C = \sum_{j\in[0,\slen{C}\sminus k)}\denote{\ket{j} \to \ket{j+k}}^{C}
\qquad
\qquad
\denote{-k}^C = \sum_{j\in[k,\slen{C})}\denote{\ket{j} \to \ket{j-k}}^{C}
\]
}

Given a quantum state $\varphi$, the addition and subtraction operations add or subtract $k$ to a basis vector; e.g., we turn $\ket{j}$ into $\ket{j+k}$ and $\ket{j-k}$, respectively.
However, we cannot predetermine the specific $j$ for a quantum state, so we use a linear sum operator to prepare the addition and subtraction operations for all possible $j\in[0,\slen{C}\sminus k)$.

The two operations above allow us to move $\ket{j+k}$ and $\ket{j-k}$ for a single data point. Now, we use these operations to define relational constraints.
We first define the $0$-flagged relational constraint, meaning all data points move in the same direction.
Here, we can pick every data-points ($\delta_j$) in $\overline{\delta}$ from the data-point set $\dfun_j\in\overline{\dfun}$.
We define the constraint as follows, with $\textcolor{spec}{\sum}$ and $\textcolor{spec}{+}$ denoting linear summation.

{\small
\[
\denote{\rset{\overline{\delta}}{0}}^{C} \triangleq \textcolor{spec}{\sum}_{k\in[0,\slen{C})} \bigcirc_{\delta\in{\overline{\delta}}} \denote{+k}^C(\cn{location}(\delta)) \,\textcolor{spec}{+} \textcolor{spec}{\sum}_{k\in[0,\slen{C})} \bigcirc_{\delta\in{\overline{\delta}}} \denote{-k}^C(\cn{location}(\delta))
\]
}

The key in the modeling is that we let all single data-points to perform the same addition or subtraction, where $\bigcirc_{\delta\in{\overline{\delta}}} \denote{+k}^C(\delta)$ refers to the sequencing operation, $\denote{+k}^C(\overline{\delta}_0) \circ ... \circ \denote{+k}^C(\overline{\delta}_{\slen{\overline{\delta}}\sminus 1})$, and $\denote{+k}^C(\delta)$ applies the Pauli-string on the qubits in the data-point $\delta$.

The modeling of the $1$-flagged ($\rset{\overline{x}}{1}$) constraint is similar to $1$-flagged one, but we ensure that every data-point adds or subtracts a different number.

{\small
\[
\denote{\rset{\overline{\delta}}{1}}^{C} \triangleq \textcolor{spec}{\sum}_{\overline{k}\in \overline{C}\wedge \slen{\overline{k}} =\slen{\overline{\delta}}} \textcolor{spec}{\sum}_{\overline{j}\in \overline{C}\wedge \slen{\overline{j}} =\slen{\overline{\delta}}} \bigcirc_{k\in \overline{k},\delta\in \overline{\delta}} \denote{\cn{compare}(k,j)}^C(\delta)
\]
}

where

{\small
\[
\cn{compare}(k,j) = + (j-k) \quad \cn{if}\, k < j
\qquad
\qquad
\cn{compare}(k,j) = - (k-j) \quad \cn{if}\, k > j
\]
}

When we pick the data-points $\overline{\delta}$, we find all $\slen{\overline{\delta}}$-length choice combinations ($\overline{k}\in \overline{C}$), where we have $\begin{pmatrix} \slen{C} \\ \slen{\overline{\delta}} \end{pmatrix}$ possible combinations.
For each combination, we let each data point ($\delta$) add or subtract the value picked at $k$.

The modeling of the swap constraint ($\rset{\overline{x}}{*}$) utilizes a similar combination concept, but we are picking different combinations of permutations; i.e., given data points $\overline{\delta}$ with the assumption of $\overline{k}\in\overline{C}$, we perform the operations based on the permuted value, as $\cn{perm}(\overline{k})$. The modeling is listed below.

{\small
\[
\denote{\rset{\overline{\delta}}{*}}^{C,} \triangleq \textcolor{spec}{\sum}_{\overline{k}\in \overline{C}\wedge \slen{\overline{k}} =\slen{\overline{\delta}}} \textcolor{spec}{\sum}_{\overline{j}\in \cn{perm}(\overline{k})} \bigcirc_{k\in \overline{k},j\in\overline{j},\delta\in \overline{\delta}} \denote{\cn{compare}(k,j)}^C(\delta)
\]
}

For example, when dealing with the clique-finding problem, to handle the constraint $\rset{x,y}{*}$, we need to consider permutations of two data points, e.g., $u$ and $v$.
Since the choice set is binary, we can write the constraint as follows.

{\small
\[
\denote{+1}(u)\circ \denote{-1}(v)+\denote{-1}(u)\circ \denote{+1}(v)
\]
}

\subsection{Compiling the Adiabatic Process to Quantum Circuits and the Theory}

After generating the Pauli-string forms for $H_i$ and $H_f$, we apply the standard approximation method for AQC in \Cref{sec:aqcbackground}, which compiles the AQC process to a sequence of Hamiltonian simulation operations, each computing $e^{-i \hat{H}(s) t}$ with $\hat{H}(s)=(1-s) H_i + s H_f$.
We compile a single Hamiltonian simulation into a quantum circuit using the optimal Trotterization method \cite{li2025verifiedcompilerquantumsimulation}.
We give the detailed circuit compilation in \Cref{appx:compilation}.
Below is the compilation correctness theorem.

\begin{theorem}[Compilation Correctness]\label{thm:compile-good}\rm
Given $\Phi$, $P$, and a desired error $r$, with data-constraint $e_U$ and relational constraint $e_S$, we compile them as $\dom{P} \vdash \Phi \gg \varphi$, $\tcompile{\Gamma}{}{e_S}{\sfun}{\hat{H}_i}$, $\tcompile{\Gamma}{}{e_U}{\sfun}{\hat{H}_f}$, and the AQC compilation as $\cn{AQC}^{r}(\hat{H}_i,\hat{H}_f)$.
The result is bounded by $r$, compared against to $P$'s dynamic semantics, as $\dabs{\cn{AQC}^{r}(\hat{H}_i,\hat{H}_f)(\varphi)-\denote{P}^{\varphi}}<r$.
\end{theorem}

\begin{proof}
   The proof has two stages. First, we show that the compilation in \Cref{fig:exp-transform} generates Pauli-string formulas with the same semantics as the \enqp constraint semantics in \Cref{sec:constrainsem}. The proof is via induction on the compilation rules in \Cref{fig:exp-transform}.
   Next, we show that AQC evolution via Hamiltonian simulation yields a unitary circuit that differs from the \enqp dynamic semantics in \Cref{def:dynsem} within a given bound. We utilize the compilation proof result in Theorem 5.8 of the QBlue compiler \cite{li2025verifiedcompilerquantumsimulation}.
\end{proof}

\subsection{Discussion: An Issue in the AQC Compilation}\label{sec:discussion}

To ensure compiler correctness, we must ensure that energy loss values in every \enqp program are positive, while the data-constraint compilation (rule \rulelab{C-EU}) in \Cref{fig:exp-transform} is applied.
Without the signs being added, the $\cn{step}^g$ function becomes monotonic decreasing within the bounded steps ($\lceil\frac{1}{\Delta t}\rceil$).
This fact is established in the proof of compilation correctness.

Because few studies have examined dynamic semantics and certified compilers, many previous works either made mistakes in their programs or did not explain why they performed such sign manipulation.
Boros \textit{et al.} \cite{Boros1991,Boros2002} did not add negative signs because they focused on developing the general concept of AQC without specifying how an AQC procedure can be compiled to quantum computers.

In \citet{farhi2000quantum}, one example, 2-SAT on a Ring, has the sign issue, as shown below.

{\small
\[
\hat{H}(s) = (1-s)\Msum_j (I - X(j)) + s \Msum_j (I- Z(j)Z(j+1))
\]
}

Note that the identity operators $I$ have no effect in compilation, so this means that we have negative signs in both the initial ($\sum_j -1\cdot X(j)$) and final ($\sum_j -1\cdot Z(j)Z(j+1)$) Hamiltonians.
This setting might work for the specific example, but it fails when we try to compile the problem for quantum computers via Hamiltonian-simulation-based compilation.

{\small
\[
\hat{H'}(s) = \Msum_{j,k} -r_{j,k} \cdot Z(j)Z(k) + \Msum_j -r_b \cdot X_j
\]
}

The work of \citet{Kadowaki_1998} and \citet{Brooke1999} develops the AQC procedure based on the classical Ising model in physics, shown above.
In the model, $r_b$ is typically negative, while a ferromagnetic model, with positive $r_{j,k}$, might be better for implementing the system on a quantum computer.
However, this description is restricted to the Ising model and assumes readers have deep knowledge of physics.
The work is also experimental for several benchmark tests, without a firm conclusion on why the signs might affect the correctness guarantee.

In industry, the Qiskit tutorial \cite{Qiskit} provides a programming pattern for negating the signs of the final Hamiltonian without explaining why users need such sign flips.

Discovering the sign issue highlights the value of developing \enqp dynamic semantics and a certified compiler.

\section{Case Study: Three-Dimensional Matching Problem}\label{sec:cases}

In this section, we show case studies of \enqp, which has a Haskell implementation. We show example programs below for solving some NP problems, and there are Haskell programs listed in \Cref{appx:cases}.

Given three finite sets $X$, $Y$, and $Z$, the three-dimensional matching problem is to find a subset $M \subseteq X \times Y \times Z$ such that no two triples in $M$ share an element.
For simplicity, we assume there is a number $n$ such that $M$ selects $n$ elements from each of the sets $X$, $Y$, and $Z$.
Additionally, we add an equational constraint across the three different sets, as follows.

{\small
\[
\forall (x,y,z) \in M \Rightarrow x \,\%\, m = y \,\%\, m = z \,\% \, m
\]
}

To program the problem in \enqp, we assume a data structure with three sets $\dfun_x$, $\dfun_y$, and $\dfun_z$; each element in the sets has a data field containing the element value and a choice field as a pair $(\zeta_1,\zeta_2)$. The first element is a Boolean value $\{\cn{Yes}, \cn{No}\}$, implementable as a single qubit, indicating whether an element is included in the set $M$. The second element is a $l$ qubit value, representing the remainder $x \,\%\, m$, $y \,\%\, m$, and $z \,\%\, m$.

We first model the \textbf{relational constraint}, as follows.

{\small
\[
\begin{array}{ll}
e_S=&
(\qbinds{(x,y)}{\dfun_x}{\qbinds{(u,v)}{\dfun_x}{\rset{x,u}{*}}})
\\[0.2em]
&\;\; + (\qbinds{(x,y)}{\dfun_y}{\qbinds{(u,v)}{\dfun_y}{\rset{x,u}{*}}})
\\[0.2em]
&\quad + (\qbinds{(x,y)}{\dfun_z}{\qbinds{(u,v)}{\dfun_z}{\rset{x,u}{*}}})
\\[0.2em]
&\quad + (\qbinds{(x,y)}{\dfun_x}{\qbinds{(u,v)}{\dfun_y}{\qbinds{(s,t)}{\dfun_z}{\rset{y,v,t}{0}}}})
\end{array}
\]
}

In the above relational constraint, the first three lines define the swap constraint for the first pair of elements in the sets $\dfun_x$, $\dfun_y$, and $\dfun_z$, where each of them constrains a MWC subgraph, similar to \Cref{fig:exmwc}.
The last line is a $0$-flagged constraint on the second pair of elements.
For every element $((x,y),(u,v),(s,t)) \in (\dfun_x,\dfun_y,\dfun_z)$, we constrain on the elements, $y$, $v$, and $t$, to ensure that they move in the same direction.
For example, if $m=4$, the three elements form the superposition state as follows.

{
\[
\frac{1}{2}(\ket{0}\ket{0}\ket{0}+\ket{1}\ket{1}\ket{1}+\ket{2}\ket{2}\ket{2}+\ket{3}\ket{3}\ket{3})
\]
}

To prepare the initial state, we first prepare three Dicke states, as described in \Cref{sec:constraintintro}, on an MWC graph similar to \Cref{fig:exmwc}. The three states are for $\dfun_x$, $\dfun_y$, and $\dfun_z$, respectively. We then compose the three MWC subgraphs; for each element with the same \cn{id} in the three subgraphs, we combine the three configurations.
We then prepare the $0$-flagged state for each of the three elements in the three sets, and we join the above Dicke state with the $0$-flagged state.

{\small
\[
\begin{array}{l@{\;}l}
e_U=&
 (\qbinds{(x,y)}{\dfun_x}{\qbinds{(u,v)}{\dfun_x}{\cif{\cn{choice}(x)=\cn{choice}(u)=\cn{Yes}\wedge\cn{data}(x)=\cn{data}(u)}{\cn{L}}{\cn{O}}}})
\\[0.2em]
&+ (\qbinds{(x,y)}{\dfun_y}{\qbinds{(u,v)}{\dfun_y}{\cif{\cn{choice}(x)=\cn{choice}(u)=\cn{Yes}\wedge\cn{data}(x)=\cn{data}(u)}{\cn{L}}{\cn{O}}}})
\\[0.2em]
&+ (\qbinds{(x,y)}{\dfun_z}{\qbinds{(u,v)}{\dfun_z}{\cif{\cn{choice}(x)=\cn{choice}(u)=\cn{Yes}\wedge\cn{data}(x)=\cn{data}(u)}{\cn{L}}{\cn{O}}}})
\\[0.2em]
& + (\qbinds{(x,y)}{\dfun_x}{\cif{\cn{data}(x)\,\%\,m\neq\cn{choice}(y)}{\cn{L}}{\cn{O}}})
\\[0.2em]
&+ (\qbinds{(x,y)}{\dfun_y}{\cif{\cn{data}(x)\,\%\,m\neq\cn{choice}(y)}{\cn{L}}{\cn{O}}})
\\[0.2em]
& + (\qbinds{(x,y)}{\dfun_z}{\cif{\cn{data}(x)\,\%\,m\neq\cn{choice}(y)}{\cn{L}}{\cn{O}}})
\end{array}
\]
}

Finally, we model the data constraint above using the generic energy values \cn{L} and \cn{O} mentioned in \Cref{sec:constraintintro}.
In $e_U$, the first three lines constrain that if two elements from the three sets are included in $M$ and share the same data, we incur an energy loss \cn{L}; otherwise, there is no loss.
The last three lines constrain the data values in the elements to satisfy the modular equation above, and we apply an energy penalty \cn{L} if they do not.

With the above modeling for the data structure, choice set, initial state, data constraint, and relational constraint, we utilize the \enqp compiler in Haskell to compile the program to produce an SQIR circuit \cite{VOQC}, and we successfully execute the program for the setting of $m=2$ and each set containing $2$ elements, via an effective simulator \cite{10.1007/978-3-032-22720-1_16}.
More case studies with Haskell programs are given in \Cref{appx:cases}.

\section{Related Work}
\label{sec:related}

We provide related work studies for AQC and quantum computation here.

\myparagraph{AQC}
The concept of quantum computation based on the adiabatic procedure, as proposed by \citet{farhi2000quantum},
and related ideas were suggested by \citet{Kadowaki_1998} and \citet{Brooke1999}.
A closely related concept is quantum annealing, a physical process that implements AQC algorithms, first proposed by \citet{Chakrabarti1981}, with many subsequent developments \cite{Ray1989,Wu1991,FINNILA1994343,Kadowaki_1998,Brooke1999}.
It incorporates algorithms and hardware designed to solve computational problems via quantum evolution towards the ground states of final Hamiltonians that encode classical optimization problems, without necessarily insisting on universality or adiabaticity.

Many NP-complete problems are framed in terms of the Ising model \cite{Barahona_1982}, a physical and mathematical model of ferromagnetism in statistical mechanics.
The idea is to frame NP-complete problems as constraint-programming problems using the Ising model and solve them in AQC or via quantum annealing \cite{Das2008,Edward2001}.
Many previous AQC-related works have been proposed to solve NP-complete problems \cite{Fu_1986,Karp1972,10.5555/314613.315014,choi2010,10.5555/2011430.2011431} by framing them as Ising models.
\citet{Lucas_2014} provides a good overview of framing different NP-problems in the Ising model.
Our clique finding algorithm is more similar to the algorithm in \citet{Wu1982}.
Note that the Ising model-framed AQC problems often go by the name QUBO (quadratic unconstrained binary
optimization) in much of the mathematical literature \cite{Boros1991,Boros2002}.
QUBO is a mathematical formalism for solving combinatorial optimization problems \cite{Kochenberge2014}.
QUBO is closely related to Ising model-based ECC problems.
D-Wave \cite{dwave2025} is a quantum computer provider that mainly uses quantum annealing to solve AQC problems with Ising-model-based Hamiltonian programs.
\enq \cite{10.1145/3747521} develops a quantum library, demonstrating how to specify Hamiltonians as data constraints that can be compiled to quantum computers.

Compared to prior work, \enqp introduces an MWC graph view of quantum states, enabling users to develop high-level programs with arbitrary data structures.
\enqp lets users define relational constraints to specify initial Hamiltonians, and develops constraint and dynamic semantics to classify the AQC procedure.
With the development of dynamic semantics, \enqp produces a certified compiler for compiling AQC problems to quantum computers, supported by theorems such as the good approximation theorem.

\myparagraph{Hamiltonian-Simulation-Based Quantum Computation.}
There are a few quantum computation prototypes allowing users to define Hamiltonians based on Pauli groups ($X$, $Y$, and $Z$),
then they are able to compile the simulation of the Hamiltonians with respect to time to quantum computers.
These works include IBM Qiskit Nature \cite{10.1145/3505636}, QuEra Bloqade \cite{bloqade2023quera},
OpenFermion \cite{mcclean2019}, InQuanto\cite{InQuanto2025}, SimuQ \cite{10.1145/3632923},
and Pasqal Pulser \cite{Silverio2022pulseropensource} developed by hardware service providers.
QuTip \cite{JOHANSSON20121760}, supporting Hamiltonian simulation in a classical computer.
These works are inspired by many classical analog compilations \cite{10.1145/3373376.3378449,10.1145/2980983.2908116}.
\citet{li2025verifiedcompilerquantumsimulation} develops a certified compiler for Hamiltonian simulation based on the Pauli-string formalism.
Most of these works support scientific computation in physics, chemistry, and biology.
The \enqp circuit compilation step follows \citet{li2025verifiedcompilerquantumsimulation}.

\myparagraph{Quantum Circuit-based Programming Languages.}
Quantum circuit-based language developments are focal.
Q\# \cite{Svore_2018}, Quilc \cite{smith2020opensource}, ScaffCC \cite{JavadiAbhari_2015}, ProjectQ \cite{Steiger_2018}, Cirq \cite{cirq_developers_2023_10247207}, Qiskit \cite{Qiskit2019} are industrial quantum circuit languages, without formal semantics, but based on a standardized denotational quantum circuit semantics.
There are formally verifying quantum circuit programs, including  \qwire~\cite{RandThesis}, \sqir~\cite{PQPC}, and \qbricks~\cite{qbricks}, quantum Hoare logic and its subsequent works \cite{qhoare,qhoreusage,10.1145/3571222}, Qafny \cite{li2024qafny}. These tools have been used to verify a range of quantum algorithms, from Grover's search to quantum phase estimation.
There are works verifying quantum circuit optimizations  (e.g., \voqc~\cite{VOQC}, CertiQ~\cite{Shi2019}), as well as verifying quantum circuit compilation procedures,
including ReVerC~\cite{reverC} and ReQWIRE~\cite{Rand2018ReQWIRERA}.

There are functional language interpretations for quantum circuits \cite{10.1007/978-3-319-89366-2_19,DIAZCARO2019104012,Selinger2013,van_Tonder_2004,Altenkirch,10.1007/978-3-642-40922-6_5,Voichick_2023,Green2013,mingsheng-ying,qml-update,qml-thesis,10.1145/3571204}, with formal semantics.
These languages mainly encode quantum circuits as functional language constructs, such as quantum pattern matching.
Silq \cite{sliqlanguage} proposed a language for automatic uncomputation, with denotational semantics to describe circuit behavior.
ZX-calculus \cite{Coecke_2011,wang2023completeness} turns elements appearing in quantum semantics into a diagram calculus for expressing unitary operations without measurement.

\section{Conclusion}

We present \enqp, a programming language and certified compiler for describing AQC problems. Most AQC problems are classical NP and P problems, such as clique finding, Hamiltonian cycle, and three-dimensional matching. Instead of low-level programming, \enqp supports high-level program descriptions based on classical constraint programming.
To help users better understand AQC problems, we develop an MWC graph-based constraint and dynamic semantics for \enqp programs and use them to certify compilation from an \enqp program to quantum computers.
While verifying the compiler, we discover a long-standing issue in Hamiltonian simulation-based AQC compilation and provide a correctness guarantee.
The good approximation theorem examines what happens when users plug in an initial state that is not an eigenstate of the relational constraint in a program.

For future work, we plan to develop software toolchains based on \enqp and explore other compilations, including classical-quantum hybrid compilation, such as QAOA \cite{Farhi2014qaoa}.

\section{Data-Availability Statement}
The software mentioned in \Cref{sec:cases,appx:cases} is available at \url{https://doi.org/10.5281/zenodo.24137515}. The artifact is produced in the described experimental setting.

\bibliography{reference}

\newpage
\appendix
\section{Compiling Hamiltonian Simulation via Trotterization}\label{appx:compilation}

\myparagraph{Performing Adiabatic Evolution.}
Once we have $H_i$ and $H_f$, we leverage AQC to find the optimum. The principle of AQC is to slowly interpolate between $H_i$ and $H_f$ over time, adiabatically guiding the quantum state from the known easy ground state to the unknown solution state. \Cref{sec:aqcbackground} introduces AQC, and we examine the intuition of performing AQC for \enqp programs.

We conceptually define a time-dependent Hamiltonian $H(s)$ that
evolves from $H(0) = H_i$ to $H(1) = H_f$ as a function of a
normalized time parameter $s \in [0,1]$. For example, a linear
schedule would use $H(s) = (1-s) H_i + s H_f$. Initially ($s=0$), the system is in the ground state of $H_i$ (the uniform superposition). As $s$ increases, the Hamiltonian gradually incorporates the problem
constraints. If the change is slow enough, the quantum adiabatic
theorem guarantees that the state of the system will remain in the
instantaneous ground state of $H(s)$. By the end ($s=1$), the system
will be in the ground state of $H_f$, which is exactly the state
encoding the optimal solution of our problem.

In practice, this continuous adiabatic evolution is realized through a
sequence of many small discrete steps. Each step applies a slight
alteration to the Hamiltonian and consequently a quantum operation on
the state. Eventually, after the full schedule, the system's state
ideally encodes the solution in its qubit amplitudes. Finally, we
measure the qubits in the computational basis. With high probability
(if the adiabatic process succeeded and if decoherence and other
errors are minimal), the measurement will collapse the system into the
basis state corresponding to the lowest energy, i.e., the optimal
solution.

Adiabatic evolution replaces the explicit search or brute-force
classical process with a physical process. As the Hamiltonian changes,
the quantum state naturally ``searches'' the landscape of
possibilities by continuously adjusting the amplitudes of basis states, concentrating more weight on low-energy (better) solutions. By the end, the amplitude is ideally concentrated on the optimal
solution. This is analogous to a landscape where the system begins as
a liquid spread out over all valleys and then slowly freezes into the lowest valley (the global minimum) due to the energy bias introduced by $H_f$. The power of this approach is that it exploits quantum
parallelism and tunneling: the state can explore many solutions
concurrently and avoid getting stuck in local minima under the right
conditions, potentially outperforming certain classical algorithms on
hard problems.

\myparagraph{Compiling the Adiabatic Process to Quantum Circuits.}

Given $H(s)$ expressed as a sum of basic operators, the next step is to turn the continuous evolution $e^{-iH(s)\Delta t}$ into a sequence of gates. \enqp uses a Trotter-Suzuki decomposition to approximate the exponential of a sum of operators. In brief, if $H = A + B$ (where $A$ and $B$ are operators that act on possibly different sets of qubits), we can approximate $e^{-i(A+B)\Delta t} \approx e^{-iA\Delta t}e^{-iB\Delta t}$ for a small time slice $\Delta t$, with the error of order $O((\Delta t)^2)$. By splitting the total evolution time $T$ into $m$ small slices ($\Delta t = T/m$) and alternating the application of $e^{-iA\Delta t}$ and $e^{-iB\Delta t}$, the accumulated error becomes negligible for large $m$. This idea generalizes to more than two terms using Suzuki’s higher-order formulas, but the principle remains: we trade a continuous time evolution for a sequence of short gate operations corresponding to each term in the Hamiltonian.

The \enqp Trotterization step is implemented based on the \qsnd compiler \cite{li2025verifiedcompilerquantumsimulation}.
We provide an example here for the following simple Ising model and present its Trotterized results.

\[
\hat{H}_{I1} = \sum_j r_h X(j) + \sum_{j} {Z(j)}\circ{Z(j\splus 1)}
    \tag{4.2.1}
  \label{eq:ising}
\]

We approximate $\eexp{\uapp{r}{\hat{H}}}$ by setting $\Pi^{N-1}_{k=0} U_k(d,\lambda,N,r)$ to be $\Pi^{d-1}_{k=0} \Pi^m \eexp{\frac{r}{m} \theta_j \hat{P}_j }$, i.e., for each Pauli string $\theta_j \hat{P}_j$, we construct an application sequence $\Pi^m \eexp{\frac{r}{m} \theta_j \hat{P}_j }$, where we repeat the simulation of $\theta_j \hat{P}_j$ for $m$ times with a scaled time slot $\frac{r}{m}$.
Each $U_k(d,\lambda,N,r)$ is defined as $\eexp{\frac{r}{m} \theta_j \hat{P}_j }$ with $N=d * m$ and $\lambda = m$.
For example, to simulate the Pauli-string-based Hamiltonian $\hat{H}_{I1}$ with a time slot $r$ and splitting number $m$, we compute the approximation of the exponent $\eexp{r{\hat{H}_{I1}}}$ as follows.

{\footnotesize
\begin{center}
$\hspace{-0.5em}
\begin{array}{l@{\;}c@{\;}l}
{\eexp{r{\hat{H}_{I1}}}}
&=& {{\Pi}^m\left(\eexp{\frac{r}{m}{(\sum_{j} \sapp{Z(j)}{Z(j\splus 1)} + r_h \sum_j X(j))}}\right)}\\
&=& {{\Pi}^m\left(\sapp{\sapp{\eexp{{{\frac{r}{m}}\sapp{Z(0)}{Z(1)}}}}{ \sapp{\eexp{{\frac{r}{m}}{\sapp{Z(1)}{Z(2)}}}}{...}}}{\sapp{\eexp{{\frac{r}{m}r_h}{X(0)}}}{\sapp{\eexp{{\frac{r}{m}r_h}{X(1)}}}{...}}} \right)}
\end{array}
$
\end{center}
}

Synthesization applies the application sequence of matrix exponentials of Pauli strings generated from Trotterization to quantum circuits, either digital or analog.
In synthesizing an application sequence $\Pi_{j} \eexp{\theta_j\hat{P}_j}$, it is enough to describe the synthesization of a single matrix exponential term $\eexp{\theta_j \hat{P}_j}$. We discuss the digital and analog-based synthesization below.

We synthesize the simulation of a Pauli string, $\eexp{\theta_j \hat{P}_j}$, to elementary quantum gates, e.g., single-qubit gates, such as \cn{H}, \cn{Rz}, \cn{Ry}, and \cn{Rx} gates, as well as two qubit controlled-not gate \cn{CX} ($\cn{CX}(x,y)$ controlling on $x$ and applying $\cn{X}$ gate on $y$).

The procedure focuses on the non-$I$ terms in a Pauli string $\hat{P}=\bigotimes_{j=0}^{n-1} \mathpzc{P}_j$. If $\hat{P}$ has only one non-$I$ term, the compilation generates a rotation gate, listed in \Cref{fig:paulisim}.
In the Bloch qubit sphere interpretation \cite{mike-and-ike}, the rotation gate $\cn{Rx}$, $\cn{Ry}$, $\cn{Rz}$ can be expressed as some Hamiltonian simulations of Pauli matrices ($\cn{Rz}(2 \theta)(j)$ applies a $Z$-axis rotation in the $j$-th qubit), which essentially suggests that the simulation of singleton Pauli $X$, $Y$, and $Z$ terms are the above phase rotation gates.
For example, the gate synthesis of $\eexp{{\frac{r}{n}}{X(j)}}$ generates an $\cn{Rx}$ gate, as $\cn{Rx}(\frac{2r}{n})(j)$.
We describe the algorithm in \Cref{fig:paulisim} for a Pauli string with more than one non-$I$ term, i.e., a $k$-local Pauli string with $k>1$.

\begin{figure}[t]
{\small
\begin{center}
$
\begin{array}{c}
{\eexp{{\theta}{X}}}=\cn{RX}(2{\theta})
=
\begin{psmallmatrix}
\cos{{\theta}} & -i\sin{{\theta}}\\
-i\sin{{\theta}} & \cos{{\theta}}
\end{psmallmatrix}
\qquad
{\eexp{{\theta}{Y}}}=\cn{RY}(2{\theta})
=
\begin{psmallmatrix}
\cos{{\theta}} & -\sin{{\theta}}\\
\sin{{\theta}} & \cos{{\theta}}
\end{psmallmatrix}
\\[0.3em]
{\eexp{{\theta}{Z}}}=\cn{RZ}(2{\theta})
=
\begin{psmallmatrix}
\eexp{{\theta}} & 0\\
0 & \cn{exp}{(i {\theta})}
\end{psmallmatrix}
\end{array}
$
\end{center}
}
  \caption{Pauli matrices simulated to rotation gates. $\eexp{\uapp{\theta}{\pau}}$ can be thought as simulating $\pau$ on time $\theta$.}
  \label{fig:paulisim}
\end{figure}

\begin{figure*}[t]
{\small
\begin{tabular}{c@{$\;\;=\;\;$}c}
\begin{minipage}{0.22\textwidth}
$
\eexp{\theta{\sapp{Z(j)}{Z(j\splus 1)}}}
$
\end{minipage}
&
\begin{minipage}{0.3\textwidth}
{
  \Qcircuit @C=0.5em @R=0.5em {
    \lstick{} & & \ctrl{1}       & \qw  & \ctrl{1}      &\qw\\
    \lstick{} & & \targ         & \gate{\cn{Rz}(2\theta)}       & \targ    & \qw
    }
}
\end{minipage}
\end{tabular}
\hspace{1em}
\begin{minipage}{0.48\textwidth}
{\small
$
\eexp{\theta ZZ}=
\begin{psmallmatrix}
\cn{exp}(\sminus i\frac{\theta}{2}) & 0 & 0 & 0\\
 0 & \cn{exp}(i\frac{\theta}{2}) & 0 & 0\\
 0 & 0 & \cn{exp}(i\frac{\theta}{2}) & 0\\
 0 & 0 & 0 & \cn{exp}(\sminus i\frac{\theta}{2})
\end{psmallmatrix}
$
}
  \end{minipage}
}
\vspace*{-0.8em}
\caption{Circuit and unitary for $ZZ$ Interaction. The left circuit applies on $j$ and $j\splus 1$ qubits.}
\label{fig:zz-inter}
\end{figure*}
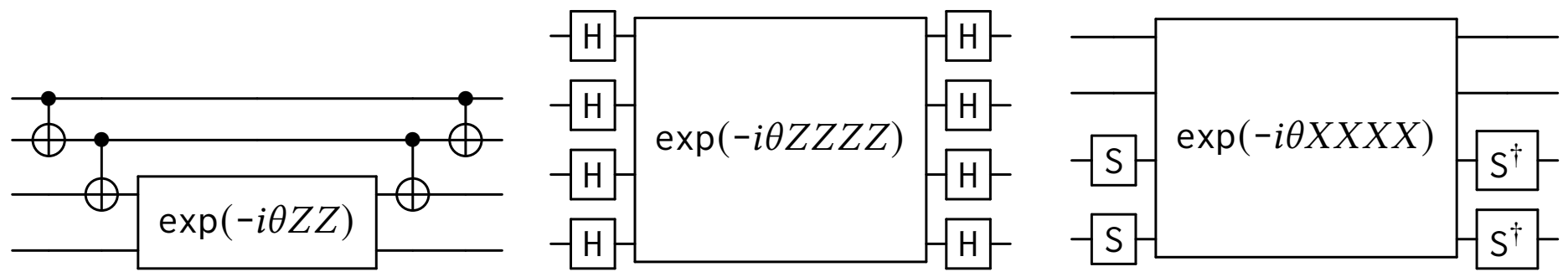

\begin{figure*}[t]
{\small
\begin{minipage}[b]{0.3\textwidth}
{\centering
{
  \Qcircuit @C=0.5em @R=0.5em {
    \lstick{} & \ctrl{1} & \qw          & \qw                            & \qw      & \ctrl{1} & \qw      \\
    \lstick{} & \targ    & \ctrl{1}     & \qw                            & \ctrl{1} & \targ & \qw \\
    \lstick{} & \qw      & \targ        & \multigate{1}{\eexp{\theta ZZ}} & \targ    & \qw   & \qw \\
    \lstick{} & \qw      & \qw          & \ghost{\eexp{\theta ZZ}}      & \qw      & \qw   & \qw
    }
}}
\subcaption{$\eexp{{\theta}{ZZZZ}}$ gate.}
\label{fig:zzzz-gate}
\end{minipage}
\begin{minipage}[b]{0.29\textwidth}
{\centering
{
  \Qcircuit @C=0.5em @R=0.5em {
    \lstick{} & \gate{\cn{H}} & \multigate{3}{\eexp{{\theta}{ZZZZ}}} & \gate{\cn{H}} & \qw   \\
    \lstick{} &\gate{\cn{H}}  & \ghost{\eexp{{\theta}{ZZZZ}}}        & \gate{\cn{H}} & \qw  \\
    \lstick{} & \gate{\cn{H}} & \ghost{\eexp{{\theta}{ZZZZ}}}        & \gate{\cn{H}} & \qw  \\
    \lstick{} & \gate{\cn{H}} & \ghost{\eexp{{\theta}{ZZZZ}}}        & \gate{\cn{H}} & \qw
    }
}}
\subcaption{$\eexp{{\theta}{XXXX}}$ gate.}
\label{fig:xxxx-gate}
\end{minipage}
\begin{minipage}[b]{0.29\textwidth}
{\centering
{
  \Qcircuit @C=0.5em @R=0.5em {
    \lstick{} & \qw & \multigate{3}{\eexp{{\theta}{XXXX}}} & \qw & \qw   \\
    \lstick{} &\qw  & \ghost{\eexp{{\theta}{XXXX}}}        & \qw & \qw  \\
    \lstick{} & \gate{\cn{S}} & \ghost{\eexp{{\theta}{XXXX}}}        & \gate{\sdag{\cn{S}}} & \qw  \\
    \lstick{} & \gate{\cn{S}} & \ghost{\eexp{{\theta}{XXXX}}}        & \gate{\sdag{\cn{S}}} & \qw
    }
}}
\subcaption{$\eexp{{\theta}{XXYY}}$ gate.}
\label{fig:xxyy-gate}
\end{minipage}
}
\caption{Synthesized circuits for some long Pauli strings.}
\label{fig:bosongates}
\end{figure*}

We first show the gate synthesization of the simple Ising model $\hat{H}_{I1}$ in \Cref{eq:ising}, which produces a sequence of $\eexp{\frac{r}{n} {Z(j)}\circ{Z(j\splus 1)}}$ and $\eexp{\frac{r}{n} r_h X(j)}$ operations (the sequence length is related to $n$). If $n$ is large enough, the operation sequence is a good approximation to $\eexp{r \hat{H}_{I1}}$.
The $\eexp{\frac{r}{n} r_h X(j)}$ operation produces a \cn{Rx} gate \cite{mike-and-ike}, while $\eexp{ \frac{r}{n} {Z(j)}\circ{Z(j\splus 1)}}$ produces a digital circuit as the left in \Cref{fig:zz-inter}, performing two controlled-not gates and an $\cn{Rz}(2\frac{r}{n})$ gate in the middle for the $j$-th and $j\splus 1$-th qubits,
i.e., $\cn{Rz}(2\theta)$ is an $z$-axis phase rotation gate.

We show an example of a long Pauli string synthesization in \Cref{fig:bosongates}.
After applying Trotterization on the $\hat{H}$, we generate a sequence of Hamiltonian simulations of Pauli strings.
Some possible Pauli string terms might be $4$-local, as $Z \otimes Z \otimes Z \otimes Z$ (abbreviated as $ZZZZ$), $X\otimes X \otimes X \otimes X$ (abbreviated as $XXXX$) and $X\otimes X \otimes Y \otimes Y$ (abbreviated as $XXYY$).
The simulation of the $ZZZZ$ term results in a quadruple $Z$ interaction in \Cref{fig:zzzz-gate}, containing a $ZZ$ interaction application in the middle, shown in \Cref{fig:zz-inter}, with the surrounding of controlled-not gates.
The digital gate synthesization of $\eexp{{\theta}{XXXX}}$ and $\eexp{{\theta}{XXYY}}$ is based on the quadruple $Z$ interaction with repairs for $X$ and $Y$ conversion, which results in the circuits in \Cref{fig:xxxx-gate,fig:xxyy-gate}.

\section{Relational Constraints in Physics}
\label{sec:physics}

Relational constraints in the programming context express a familiar physical idea: restricting evolution to a chosen subspace by allowing only transitions that preserve it. In quantum annealing, these transitions define a \emph{driver Hamiltonian}; in gate-based optimization, they define a \emph{mixer}. When the constraints fix the value of a \emph{conserved quantity}, the subspace is a \emph{symmetry sector}. Table~\ref{tab:physics-dictionary} summarizes these connections. This appendix reviews the physical foundations of relational constraints, the guarantees and limitations of existing constructions, and the tasks they leave to the programmer.

\begin{table}[t]
\caption{\enqp{} terms and their names in physics.}
\label{tab:physics-dictionary}
\small
\begin{tabular}{@{}p{0.42\textwidth}p{0.52\textwidth}@{}}
\toprule
\enqp{} & Physics \tabularnewline
\midrule
cost function; its minimum & Hamiltonian; ground state \tabularnewline
data constraint $\hat H_f$ & problem Hamiltonian; phase separator (QAOA) \tabularnewline
relational constraint $\hat H_i$ & driver Hamiltonian (annealing); mixer (QAOA) \tabularnewline
quantity a relational constraint preserves & conserved quantity; symmetry \tabularnewline
candidates (nodes of the MWC graph) & symmetry sector; feasible subspace \tabularnewline
swap of a chosen and an unchosen vertex & XY exchange; hard-core boson hopping; Kawasaki move \tabularnewline
uniform start over the $k$-subsets & Dicke state \tabularnewline
penalty $(k-\sum_v x_v)^2$ & energy penalty; grand-canonical control of particle number \tabularnewline
\bottomrule
\end{tabular}
\end{table}

\paragraph{Conservation laws}
When the Hamiltonian commutes with an observable throughout the evolution, it
preserves each eigenspace of that observable. In our clique
example, the conserved
observable is particle number: each vertex is represented by a mode that is either
empty or occupied, and selecting exactly $k$ vertices corresponds to the
$k$-particle subspace. This restriction to at most one particle per mode occurs
for hard-core bosons, fermions, and hard-core anyons, although their exchange
statistics differ~\citep{Somma2003}. Our relational swap corresponds to hard-core
boson hopping, \(\theta = 2\pi \): it transfers a particle from an occupied mode to an empty one while
preserving the total particle number. For $n$ such modes, the Hilbert space
decomposes into fixed-particle-number sectors:
\[
  \mathcal H = \bigoplus_{k=0}^{n} \mathcal H_k,
  \qquad
  \mathcal H_k = \ker(\hat N-kI),
  \qquad
  \hat N = \sum_{v=1}^{n}|1\rangle\langle 1|_v.
\]
Here $\hat N$ counts occupied modes, and $\mathcal H_k$ is the subspace with
exactly $k$ particles.
An example of particle-number-preserving hopping appears in the resonant impurity
scattering simulation reported by \citet{Ortiz2002}. Conservation can be used to
reduce the simulation space: \citet{Graves2025} restricts an anyonic
Aharonov--Bohm Hamiltonian to fixed-particle-number subspaces while retaining their
exact dynamics. In \enqp{}, the diagonal data Hamiltonian also preserves particle
number, so evolution initialized in the $k$-particle subspace remains there. The
Dicke state~\citep{Dicke1954} used by the program is the uniform superposition of
all occupation patterns with exactly $k$ particles. This enforces cardinality
through conservation, whereas an energy penalty makes violations costly without
itself forbidding transitions into them. The same principle gives
invariant subspaces in spin systems: the zero-field Ising transfer matrix commutes
with global spin reversal and therefore preserves sectors of even and odd
spin-reversal parity~\citep{Baxter1982}.

\paragraph{Constraint-preserving drivers and mixers}
\citet{10.5555/2011430.2011431} searched for cliques exactly this way. Their initial cost function
summed swaps over all pairs of vertices, so the run stayed among the
subsets of the target size, and it began in the uniform superposition
of those subsets: the program of this
section. \citet{Hen2016constrained} turned the construction into a
general method,
\emph{constrained quantum annealing}.
Later work expanded constraints~\citep{HenSarandy2016,Kudo2018coloring} and
general linear equalities~\citep{Leipold2022linear,Leipold2024imposing}. QAOA mixers provide a gate-based counterpart to constraint-preserving drivers.
In the quantum alternating operator ansatz~\citep{Hadfield2019qaoa}, mixer
unitaries preserve the feasible subspace, while cost unitaries encode the
optimization objective. This parallels \enqp{}'s separation of relational from
data constraints. For example, XY mixers preserve Hamming weight and can therefore
enforce one-hot and cardinality constraints~\citep{Wang2020xy}. The original EnQ implementation used a QAOA style mixer in its gate based circuit
backend~\citep{10.1145/3747521}.

\section{Additional Case Studies}\label{appx:cases}

In this section, we show three case studies for the utility of \enqp.

\subsection{The Compilation Pipeline of Hamiltonian Cycle}\label{appx:numberpartitioning}

\begin{figure}[ht]
\centering

\begin{minted}[
    fontsize=\small,
    breaklines,
    linenos,
    numbersep=5pt,
    xleftmargin=1em
]{haskell}
hamiltonianCycle :: [(Int, Int)] -> Program [] Int Int Int
hamiltonianCycle edges =
  let 
    cities = getNodes edges
    numNodes = length cities
  in 
  Program
    { dataProgram = DataProgram
        { choices = [0 .. numNodes-1]
        , struct = cities
        , view = 2
        , constraints = \case
            [(cityA, rankA), (cityB, rankB)] ->
               let
                 diff = abs (rankA - rankB)
                 isAdjacentRank = diff == 1 || diff == (numNodes - 1)
                 
                 connectivityPenalty =
                   if isAdjacentRank
                   then if (cityA, cityB) `elem` edges || (cityB, cityA) `elem` edges
                        then 0 
                        else 50
                   else 0
               in connectivityPenalty
            _ -> error "hamiltonianCycle: Expected exactly two choices"
        , choiceLayout = FlatEncoding
        }
    , relationalProgram = [mkRel Swap 2 ((-1.0) :+ 0.0)]
    , initialState = Permutation
    }
\end{minted}

\caption{The Hamiltonian Cycle implementation in \enqp}
\label{fig:hamiltonian}
\end{figure}

The Hamiltonian Cycle problem involves finding a closed loop through a given graph
$G = (V, E)$ that visits every vertex exactly once. Formulating Hamiltonian Cycle
within our Energy Constraint Computation (ECC) framework requires a careful
encoding of both the vertex permutation and the graph's adjacency constraints.

As shown in Figure~\ref{fig:hamiltonian}, we model the cycle using a rank-based positional encoding. For a graph with $N$ vertices, the \mintinline{haskell}{struct} defines the set of vertices (cities), and the available \mintinline{haskell}{choices} correspond to the  ranks $0$ through $N-1$. The \mintinline{haskell}{constraints} function then evaluates whether an edge exists between the chosen vertex at rank $i$ and the vertex at rank $i+1$. If $(v_i, v_{i+1}) \notin E$, an energy penalty is applied. To ensure the path forms a closed loop, the \mintinline{haskell}{constraints} function also wraps around to evaluate the adjacency between the final step and the first step.

The data constraints define the problem's energy landscape. By setting
\hask{view} to $2$, the compiler evaluates pairwise relationships between the
cities. For a given pair $(\texttt{cityA}, \texttt{rankA})$ and
$(\texttt{cityB}, \texttt{rankB})$, the function calculates the absolute
difference in their assigned ranks. We evaluate our compilation using a
4-node cycle graph ($C_4$), where $V = \{0,1,2,3\}$ and the valid edges form
the perimeter of a square. If the difference is exactly $1$, or $3$
(enforcing the wrap-around condition to close the 4-step loop), the cities are
deemed adjacent in the sequence. If these adjacent cities do not share a valid
edge in the underlying graph, that is, if the sequence attempts to traverse
the missing diagonal edges $(0,2)$ or $(1,3)$, a connectivity penalty of
$50$ is applied.

\subsubsection{Relational Constraints: The Initial Hamiltonian}

To avoid exploring a massive subspace of invalid configurations (e.g., assigning
the same rank to multiple cities), \enqp restricts the Many-World
Configuration (MWC) graph to search only over valid permutations. This is
achieved by initializing the system in a valid permutation state and driving
the evolution with a relational constraint that strictly enforces rank-swapping
between nodes. As established in Section~\ref{sec:initham}, the compilation of the relational swap
constraint ($\bar{x}^{\sim *}$) into a quantum circuit proceeds in two phases.

\paragraph{Implementing Addition and Subtraction Operators.}

The compiler first constructs the raising operators
$\llbracket +k \rrbracket^{C}$ and lowering operators
$\llbracket -k \rrbracket^{C}$ over the binary representations of the sequence
ranks. For our four ranks,
$C=\{0,1,2,3\}$, each rank is encoded using two qubits as
$|00\rangle$, $|01\rangle$, $|10\rangle$, and $|11\rangle$.

Suppose the evolution requires shifting a rank by $+2$. The corresponding
raising operator induces the following state transitions:

\[
|00\rangle \xrightarrow{\llbracket +2 \rrbracket^{C}} |10\rangle,
\]

\[
|01\rangle \xrightarrow{\llbracket +2 \rrbracket^{C}} |11\rangle,
\]

\[
|10\rangle,\;|11\rangle
\xrightarrow{\llbracket +2 \rrbracket^{C}}
0,
\]

where the latter two states are annihilated because they cannot be shifted by
two within the valid rank space.

At the Pauli level, the transition
$|00\rangle \rightarrow |10\rangle$
requires applying the creation operator
$a^\dagger=\frac{1}{2}(X-iY)$
to the first qubit while projecting the second qubit onto
$|0\rangle\langle0|=\frac{1}{2}(I+Z)$.
The resulting tensor-product representation is:

\[
\frac{1}{2}(X-iY)\otimes\frac{1}{2}(I+Z).
\]

Similarly, the transition
$|01\rangle \rightarrow |11\rangle$
keeps the second qubit in the
$|1\rangle$ state by applying the projector
$|1\rangle\langle1|=\frac{1}{2}(I-Z)$
giving:

\[
\frac{1}{2}(X-iY)\otimes\frac{1}{2}(I-Z).
\]

Therefore, the complete raising operator
$\llbracket +2 \rrbracket^{C}$
over the rank subspace is obtained by summing the two transitions:

\[
\llbracket +2 \rrbracket^{C}
=
\left(
\frac{1}{2}(X-iY)\otimes\frac{1}{2}(I+Z)
\right)
+
\left(
\frac{1}{2}(X-iY)\otimes\frac{1}{2}(I-Z)
\right)
\]

\paragraph{Implementing the Relational Swap.}

With these base operators constructed, \enqp pre-generates the valid
permutation set and examines the relations required to implement swaps for each
transition in the Many-World Configuration (MWC) graph. For example, consider our $C_4$ graph again. If city $0$ currently holds rank $0$
($|00\rangle$) and city $2$ holds rank $2$ ($|10\rangle$), swapping their
sequence assignments requires applying a $+2$ shift to city $0$ and a $-2$
shift to city $2$. The compiler therefore constructs the joint transition:

\[
\llbracket -2 \rrbracket(\text{city }2)
\circ
\llbracket +2 \rrbracket(\text{city }0).
\]

Likewise, the reverse transition is represented by

\[
\llbracket -2 \rrbracket(\text{city }0)
\circ
\llbracket +2 \rrbracket(\text{city }2).
\]

By pre-generating the valid permutation set and applying the generalized
compare summations described in Section~\ref{sec:initham} across all node pairs,
\enqp constructs the complete initial Hamiltonian ($H_i$). This
guarantees that the initial Hamiltonian rigorously preserves the
permutation subspace without leaking amplitude into invalid configurations. As
the adiabatic evolution proceeds, the system explores only the valid paths
defined by the edges of the MWC graph, significantly accelerating convergence
toward the optimal Hamiltonian cycle encoded by the data constraints.

\subsubsection{Execution and Evolution}

Execution begins by initializing the quantum register into the permutation
subspace using a Quantum Fisher--Yates Shuffle, which prepares an equal
superposition over the $24$ valid permutations of our $C_4$ graph. Driven by the relational constraints, the Trotterized adiabatic evolution swaps
ranks between cities while strictly traversing the valid edges of the MWC
graph. This strict confinement allows the system to effectively steer the quantum state toward the optimal Hamiltonian cycle.

\subsection{Traveling Salesperson}\label{appx:exactcover}

The Traveling Salesperson Problem (TSP) is a natural extension of the
Hamiltonian Cycle problem. Given a graph with weighted edges representing
distances (or costs) between cities, the objective is to find a continuous
closed loop that visits every city exactly once while minimizing the total
accumulated travel cost.

Within our Energy Constraint Computation (ECC) framework, TSP utilizes the same
positional encoding as the Hamiltonian Cycle. The data structure,
\hask{struct}, represents the set of cities, while \hask{choices} represent the
sequential ranks ($0$ to $N-1$) assigned to each city. The compiler evaluates
pairs of cities. For a given pair
$(\texttt{cityA}, \texttt{rankA})$ and
$(\texttt{cityB}, \texttt{rankB})$, the function computes the absolute
difference between their assigned ranks to determine adjacency, including the
$N-1$ wrap-around condition. The key distinction is that TSP directly maps the continuous edge weights to
the energy eigenvalues of the problem Hamiltonian, whereas the Hamiltonian
Cycle problem applies a fixed penalty for missing edges.

\subsubsection{Relational Constraints and Evolution}

Because the structural requirement of TSP, namely visiting every city exactly
once, is identical to that of the Hamiltonian Cycle problem, the search space
and relational constraints established in
Section~\ref{sec:initham} can be reused without modification. The quantum register is initialized in the valid permutation subspace using the
Quantum Fisher--Yates Shuffle. The adiabatic evolution then employs the same
relational constraint that strictly enforces rank-swapping between nodes. Consequently, the
compiler reuses the same addition and subtraction operators together with the
multi-node swap operations to construct the initial Hamiltonian. As the adiabatic evolution proceeds, the system traverses only the valid edges
of the Many-World Configuration (MWC) graph. The resulting energy landscape
naturally guides the quantum state toward the sequence that minimizes the
accumulated data constraints, driving the system toward the optimal Traveling
Salesperson tour.

\subsection{Equational Three-Dimensional Matching}\label{appx:exactcover}

Given three finite sets $X$, $Y$, and $Z$, the classic Three-Dimensional Matching (3DM) problem seeks a subset $M \subseteq X \times Y \times Z$ such that no two triples share an element. For this case study, we assume the subset $M$ selects exactly $n$ elements from each set. Furthermore, we expand the classical formulation by introducing a global equational constraint across the chosen elements:
\[
\forall(x, y, z) \in M \implies x \% m = y \% m = z \% m
\]

\enqp allows us to model this complex variant seamlessly by structuring the elements and choices as follows:

\begin{minted}[fontsize=\small, breaklines, linenos, xleftmargin=1em]{haskell}
-- | Which partite set an equational-3DM element belongs to.
data MatchSet = SetX | SetY | SetZ deriving (Show, Eq, Ord)

-- | Element: (set tag, numeric value).
type MatchElem = (MatchSet, Int)

-- | Choice: (in matching?, claimed remainder mod m).
type MatchChoice = (Bool, Int)
\end{minted}

We structure elements into disjoint sets ($\mathcal{D}_x, \mathcal{D}_y, \mathcal{D}_z$) and encode each element's \mintinline{haskell}{MatchChoice} as a tuple: a Boolean for inclusion in $M$, and an integer representing its remainder modulo $m$.

\subsubsection{Data Constraints: The Problem Hamiltonian}
The data constraints evaluate the pairwise consistency of the elements to define the problem's energy landscape.

\begin{minted}[fontsize=\small, breaklines, linenos, firstnumber=10, xleftmargin=1em]{haskell}
    , dataProgram = DataProgram
        { choices = choiceList
        , struct = elems
        , view = 2
        , constraints = \case
            [((setA, valA), (yesA, remA)), ((setB, valB), (yesB, remB))] ->
              let
                uniq =
                  if setA == setB && yesA && yesB && valA == valB
                  then penalty
                  else 0
                remAOK = if valA `mod` m == remA then 0 else penalty
                remBOK = if valB `mod` m == remB then 0 else penalty
              in uniq + remAOK + remBOK
            _ -> error "equational3DM: Expected exactly two choices"
        , choiceLayout = InclusionRemainder m
        }
\end{minted}

By setting the \mintinline{haskell}{view} to $2$, the compiler evaluates every pair of candidates. The constraint imposes an energy penalty if two elements belong to the same partite set (\texttt{setA == setB}), are both chosen for inclusion (\texttt{yesA \&\& yesB}), and correspond to the same underlying value (\texttt{valA == valB}). This natively prevents invalid overlaps in $M$. Additionally, the constraint calculates the actual arithmetic remainder of the data (\texttt{valA `mod` m}) and applies a penalty if it mismatches the dynamically chosen remainder state (\texttt{remA}).

By setting \mintinline{haskell}{view} to~$2$, the compiler evaluates every pair of candidates.
The constraint imposes an energy penalty if two elements belong to the same partite set (\texttt{setA == setB}), are both chosen for inclusion (\texttt{yesA \&\& yesB}), and correspond to the same numeric value (\texttt{valA == valB}), guarding against
duplicate labels within a partite set.
(Selecting exactly~$n$ elements from each of~$X$,~$Y$, and~$Z$ is enforced
separately by the Dicke initial state of weight~$n$ per set together with
the Swap mixer on inclusion bits.)
A second term compares each site's value modulo~$m$
(\texttt{valA `mod` m}) with its claimed remainder register
(\texttt{remA}) and penalizes mismatches, so feasible configurations are
those whose selected values all leave the same residue modulo \(m\).

\subsubsection{Relational Constraints and State Preparation}
The true expressiveness of \enqp is demonstrated in how it restricts the massive search space to only valid subset sizes and consistent remainders. The relational constraint ($e_S$) for Equational 3DM is heavily stratified:

\begin{minted}[fontsize=\small, breaklines, linenos, firstnumber=27, xleftmargin=1em]{haskell}
    , relationalProgram =
        [ swapOn xIxs
        , swapOn yIxs
        , swapOn zIxs
        , RelationalProgram
            { relationalView = 3
            , rule = ZeroFlag
            , theta = (-1.0) :+ 0.0
            , siteMask = AllSites
            , fieldMask = RemainderField
            }
        ]
    , initialState = Equational3DM n m setSize
\end{minted}

We apply independent swap rules (\mintinline{haskell}{swapOn}) exclusively on the inclusion bits within $X$, $Y$, and $Z$. This ensures that if an element is added to $M$, another from the exact same set must be removed, rigidly preserving the constraint that exactly $n$ elements are selected per set. Simultaneously, we apply a $0$-flagged relational constraint (\mintinline{haskell}{ZeroFlag}) across the remainder fields of all elements. This dictates that the chosen remainder modulo $m$ must transition uniformly across every element simultaneously, enforcing the equational constraint inherently. To initialize this constrained subspace, the compiler constructs an initial quantum state by first preparing three independent Dicke states on the inclusion qubits of sets $X, Y$, and $Z$. Concurrently, the system initializes the remainder fields into a globally entangled, uniform $0$-flagged superposition (e.g., $\frac{1}{\sqrt{m}}\sum_r |r\rangle \dots |r\rangle$). This synchronized state perfectly aligns with the ground state of the $0$-flagged initial Hamiltonian, embedding the equational constraint from the outset.

\subsubsection{Execution and Evolution}
Driven by the relational constraints, the Trotterized adiabatic evolution independently swaps element inclusions within their respective partite boundaries while globally shifting the entangled remainder value in lockstep. By traversing only the valid edges of the MWC graph, this strict subspace confinement allows the energy landscape to effectively steer the adiabatic evolution toward the optimal set of valid triplets.

\end{document}